\documentclass{article}

\pdfoutput=1

\usepackage[T1]{fontenc}
\usepackage[utf8]{inputenc}
\usepackage{lmodern}
\usepackage{microtype}

\usepackage[
  letterpaper,
  top=1in, bottom=1in,
  left=0.8in, right=0.8in
]{geometry}

\usepackage{amsmath,amssymb,amsthm}
\usepackage{bm}
\usepackage{booktabs}
\usepackage{array}
\usepackage[table]{xcolor}
\usepackage{graphicx}
\usepackage{float}
\usepackage{authblk}
\usepackage{natbib}

\usepackage{subcaption}

\usepackage[colorlinks=true,linkcolor=blue,citecolor=blue,urlcolor=blue]{hyperref}

\title{Deep description of static and dynamic network ties in Honduran villages}

\author[1]{Marios Papamichalis}
\author[1]{Nikolaos Nakis}
\author[1,2,3]{Nicholas A. Christakis}

\affil[1]{Human Nature Lab, Yale University}
\affil[2]{Department of Statistics and Data Science, Yale University}
\affil[3]{Department of Sociology, Yale University}

\affil[ ]{\texttt{marios.papamichalis@yale.edu}; \texttt{nicolaos.nakis@gmail.com}; \texttt{nicholas.christakis@yale.edu}}

\date{}

\begin{document}
\maketitle
\begin{abstract}
We examine static and dynamic social network structure in 176 villages within the Copán Department of Honduras across two data waves (2016, 2019), using detailed data on \emph{multiplex} networks for 20,232 individuals enrolled in a longitudinal survey. These networks capture friendship, health advice, financial help, and adversarial relationships, allowing us to show how cooperation and conflict jointly shape social structure. Using node-level network measures derived from near-census sociocentric village networks, we leverage mixed-effects zero-inflated negative binomial models to assess the influence of individual attributes, such as gender, marital status, education, religion, and indigenous status, and of village characteristics, on the dynamics of social networks over time. We complement these node-level models with dyadic assortativity (odds-ratio-based homophily) and community-level measures to describe how sorting by key attributes differs across network types and between waves. Our results demonstrate significant assortativity based on gender and religion, particularly within health and financial networks. Across networks, gender and religion exhibit the most consistent assortative mixing. Additionally, community-level assortativity metrics indicate that educational and financial factors increasingly influence social ties over time. Our findings provide insights into how personal attributes and community dynamics interact to shape network formation and socio-economic relationships in rural settings over time.
\end{abstract}


\section{Introduction}

Social networks shape individual behaviors, access to resources, and overall well-being \citep{smith2008social}. They are associated with information flow, distribution of support, and the maintenance of social norms within communities. Investigating the factors that determine the formation and maintenance of social networks sheds light on how communities are bound together. This understanding is particularly vital in rural settings, where resources are often scarce, and social ties significantly impact individuals' access to support and opportunities.\\

Here, we investigate the determinants of health, friendship, financial, and adversarial relationships among 20,232 individuals in 176 rural Honduran villages. By focusing on these diverse types of relationships, we aim to capture a holistic view of social network dynamics within these communities. Health advice relationships are crucial for understanding how individuals seek and provide care, while friendship ties reveal patterns of social support and companionship. Financial relationships shed light on economic exchanges and resource sharing. Adversarial relationships highlight conflict and social tensions. By examining these relationships, we identify the socio-economic and demographic factors that are associated with network formation and maintenance.\\

Village social life in low‑infrastructure settings is multiplex: the same households exchange health advice, friendship, and financial help while also experiencing conflict. These ties shape how information moves, who receives support, and where disputes arise. Decades of work show that network structure can be leveraged to deliver programs and trace behavior change, from microfinance diffusion to health messaging and peer‑influence experiments \citep{christakis2010social,banerjee2013diffusion,centola2010spread,centola2011experimental,kim2015social,hunter2015hidden}. At the same time, homophily and community structure can segment populations and limit reach. In single‑layer village networks, communities are often ``connected yet segregated,'' with sharp alignment to social attributes such as caste or sex \citep{montes2018connected,ghasemian2024structure}. \\

More specifically, empirical analyses of rural village networks across the Global South reveal dense, highly clustered structures that are nonetheless segregated along social lines, with substantial heterogeneity in individual positions and tie types that shapes diffusion and program reach \cite{perkins2015social}. In 75 Karnataka villages, networks aggregated from 12 relationship types featured large connected components (typically \(\sim\)98\% of nodes) and high clustering (\(\sim 0.6\)), yet strong dyad-level assortativity by caste (significant in 98\% of villages; same-caste tie odds ratios \(\approx 1.1\)–18.6) and by sex, and community partitions that aligned most with caste (normalized mutual information \(\approx 0.39\)), while sex explained little of modular structure; \citep{montes2018connected}. These patterns confirm homophily and triadic closure as organizing mechanisms in rural networks and help explain why diffusion often proceeds within identity-defined modules unless bridges are present \citep{newman2003mixing,henry2011emergence,fortunato2010community,blondel2008fast}.\\

Related empirical work shows that network structure, rather than geography or simple demographics, predicts spread of innovations and behaviors in villages: microfinance participation in India followed network paths and was anticipated by centrality-based measures \citep{banerjee2013diffusion}; experimental and field studies demonstrate that homophily and clustered topology modulate adoption dynamics \citep{centola2010spread,centola2011experimental}, and that network-informed targeting, e.g., seeding nominated friends to exploit the friendship paradox, can outperform naive strategies in increasing population-level uptake \citep{kim2015social,hunter2015hidden,airoldi2024induction}. \\

Beyond single-layer representations, the literature emphasizes multiplexity, distinct layers such as kinship, advice, and resource exchange that are not redundant and can differentially mediate contagion, implying that aggregating ties may obscure critical pathways for association \citep{porter2009communities,kumpula2007emergence,airoldi2024induction}. At the same time, identity-based segregation (e.g., caste, religion, gender) and traditional institutions shape access to information and opportunity, generating village-to-village and within-village heterogeneity in network roles (hubs, brokers) and exposure to external ideas \citep{munshi2006traditional,sanyal2015group,berreman1972race,patil2014caste}.\\

Methodologically, studies have leveraged a community detection and modularity \citep{girvan2002community,fortunato2010community,blondel2008fast}, assortativity metrics \citep{newman2003mixing,noldus2015assortativity}, and statistical network models \citep{hoff2002latent,wang1987stochastic,robins2007introduction}, while design-oriented work has highlighted that ignoring homophily can bias seeding evaluations and that connectivity-informed designs can strengthen cluster-randomized trials \citep{aral2013engineering,harling2017leveraging}.\\

Taken together, the evidence indicates that rural networks are cohesive yet partitioned; interventions should therefore seed multiple communities, engage both hubs and brokers, and use layer-appropriate channels to traverse homophilous boundaries and achieve equitable diffusion \citep{montes2018connected,banerjee2013diffusion,kim2015social}. Yet, despite this foundation, we lack an integrated account of \emph{multiplex} rural structure that (i) quantifies how cooperative layers, \emph{friendship}, \emph{financial help}, and \emph{health advice}, couple at the person and place levels, and (ii) positions negative ties within that cooperative scaffold. We address these gaps using a two‑wave, near‑census multiplex of $V{=}176$ Honduran villages (2016, 2019), measuring four undirected layers per village: \emph{health}, \emph{friendship}, \emph{financial}, and \emph{adversarial}. Our empirical framework spans nodes, dyads, communities, and villages. We estimate multivariate mixed models for degree (with village/wave heterogeneity and a Mundlak decomposition); test assortativity with degree‑aware label permutations; and detect communities on the positive union and relate them to attributes using normalized mutual information (NMI) and within‑village regularized multinomial models.\\

Several robust regularities emerge. \emph{First}, cooperative layers co‑move strongly at both the village and individual levels yet show little dyadic redundancy: places dense in friendship are also dense in health and financial help, but the same pairs seldom carry more than one cooperative tie. \emph{Second}, Schooling and age are associated with more cooperative and fewer adversarial ties; men concentrate in financial exchange while women anchor friendship/health; marriage broadens engagement; income sufficiency shifts ties toward financial help and lowers conflict; major religious denominations show mild cooperative advantages; indigenous identity is linked to slightly higher degrees across layers; and place correlates are weak on average. \emph{Third}, \textit{assortativity and degree‑aware permutations} show pervasive same‑gender and same‑religion mixing across cooperative layers (with little cross‑gender dissortativity), while other attributes (age, education, income, indigenous status) are modest and heterogeneous once degree is controlled. \emph{Fourth}, multiplex communities have moderate--to--high modularity and align most closely with friendship; single attributes convey limited information about community labels, though gender, age, marital status, and education remain predictive in within‑village multinomial fits. In sum, we offer a unified portrait of rural multiplex structure that links segregation, cross‑layer coupling, and methodologically, we combine multivariate GLMMs, degree‑aware nulls and signed‑network diagnostics.\\

The paper is organized as follows. Section~2 details the setting, measurement, four layers (friendship, financial, health, adversarial), covariates, and the construction of the positive union and signed networks, and reports descriptive network characteristics by layer. Section~3 covers: village/individual co‑movement vs dyadic overlap; the traits of individuals who have more connections in both cooperative and conflict layers, with village/respondent heterogeneity and a Mundlak decomposition; dyad‑level assortativity with degree‑aware permutation tests; communities and their alignment with attributes. Section~4 concludes.

\section{Data}

This study was conducted in the Copan Department of Honduras, covering a geographic area of over 200 square miles. We began with 176 villages with 32,800 health-related eligible participants, of which 93\% ($N$=30,422) consented to be included in our census. From this population, we enrolled $n=20{,}232$ individuals for participation in the two waves used in the present study (labeled wave 1 and wave 3). These participants expressed willingness to participate in surveys, undergo randomization, and engage in a potential 22-month intervention. They also committed to ongoing participation in annual or biannual surveys.\\

Due to the geographic isolation of the Copan villages, social networks were primarily intravillage. To capture the complexity of these networks, we used publicly available software, detailed in \cite{lungeanu2021using}, which facilitated the mapping of these intricate social connections. Table \ref{tab:name_generator} presents the questionnaire items (“name generators”) used to construct four distinct social networks, health, financial, friendship, and adversarial, based on self-reported interpersonal connections. All networks are undirected, with at most one edge between any pair of nodes, derived from directed survey responses (by symmetrizing nominations). Table~\ref{tab:variables_summary} summarizes the individual- and village-level variables used throughout the analyses. Further details regarding the statistical analysis of each characteristic and additional data related information are presented in the Appendix~\ref{app:timeline}.



\begin{table}[!ht]
\centering
\setlength{\tabcolsep}{10pt}
\renewcommand{\arraystretch}{1.20}

\begin{tabular}{>{\raggedright\arraybackslash}p{2.6cm} >{\raggedright\arraybackslash}p{10.8cm}}
\rowcolor{blue!20}
\textbf{Connection} & \textbf{User} \\

\rowcolor{blue!10}
Health & Who would you ask for advice about health-related matters? \\

\rowcolor{blue!10}
Health & Who comes to you for health advice? \\

\rowcolor{blue!10}
Friendship & Who do you trust to talk to about something personal or private? \\

\rowcolor{blue!10}
Friendship & With whom do you spend free time? \\

\rowcolor{blue!10}
Friendship & Besides your partner, parents, or siblings, who do you consider to be your closest friends? \\

\rowcolor{blue!10}
Financial & Who would you feel comfortable asking to borrow 200 lempiras from if you needed them for the day? \\

\rowcolor{blue!10}
Financial & Who do you think would be comfortable asking you to borrow 200 lempiras for the day?\\

\rowcolor{blue!10}
Adversarial & Who are the people in this village with whom you do not get along well?\\
\end{tabular}

\vspace{2pt}
\caption{Questionnaire with Connection Type (``Name Generators'')}
\label{tab:name_generator}
\end{table}

\begin{table}[H]
\centering
\caption{Variables used in the analyses.}
\label{tab:variables_summary}
\small
\begin{tabular}{ll}
\toprule
Category & Variables \\
\midrule
\textbf{Individual} &
Gender; Age; Education (years); Marital status; Religion; \\
& Income sufficiency (self‑reported); Indigenous identity \\
\addlinespace
\textbf{Village / spatial} &
Village size (surveyed adults); Travel time to health center; \\
& Travel time to main road; Travel time to maternity clinic; \\
& Household geocoordinates (for pairwise distances) \\
\addlinespace
\textbf{Networks (per village, per wave)} &
Four undirected layers: Friendship, Financial, Health, Adversarial; \\
\bottomrule
\end{tabular}
\end{table}

\subsection*{Network characteristics by layer}

\subsection*{Friendship networks}
We summarize the structure of the friendship layer in Table~\ref{tab:friendship_w1w3}. 
Across waves, networks are moderate in size (median \(N=80\); range 16--370) and relatively dense for village social networks (median \(\delta\approx0.06\) in Wave~1 and \(0.058\) in Wave~3) with noticeable clustering (median \(\langle C\rangle=0.29\) and \(0.32\), respectively). 
Fragmentation is limited (median of 5 vs.\ 6 components). 
For analysis we focus on the largest connected components (LCCs), which contain a median of \(95\%\) (Wave~1) and \(94\%\) (Wave~3) of nodes and essentially all ties (median \(m/M=100\%\) in both waves). 
Median mean degree in the LCC is very close to that of the full network (about \(+4\%\) in Wave~1 and \(+6\%\) in Wave~3), and median clustering is virtually unchanged, mirroring the pattern highlighted by \cite{montes2018connected}.

\begin{table}[H]
\centering
\footnotesize
\caption{Friendship layer (Wave~1 and Wave~3): characteristics of full networks and LCCs.}
\label{tab:friendship_w1w3}
\begin{tabular}{lccc ccc ccc ccc}
\toprule
& \multicolumn{6}{c}{\textbf{Wave 1}} & \multicolumn{6}{c}{\textbf{Wave 3}} \\
\cmidrule(lr){2-7}\cmidrule(lr){8-13}
& \multicolumn{3}{c}{Networks} & \multicolumn{3}{c}{LCC} & \multicolumn{3}{c}{Networks} & \multicolumn{3}{c}{LCC} \\
\cmidrule(lr){2-4}\cmidrule(lr){5-7}\cmidrule(lr){8-10}\cmidrule(lr){11-13}
\textbf{Metric} & Min & Med & Max & Min & Med & Max & Min & Med & Max & Min & Med & Max \\
\midrule
$N$                     & 16 & 80  & 370  & 15 & 75 & 359 & 16 & 80  & 370 & 15 & 74 & 358 \\
$M$                     & 23 & 202 & 1066 & 23 & 202 & 1065 & 34 & 180 & 805 & 34 & 180 & 802 \\
$\delta$                & 0.011 & 0.062 & 0.206 & 0.013 & 0.071 & 0.219 & 0.011 & 0.058 & 0.283 & 0.012 & 0.065 & 0.324 \\
$\langle k\rangle$      & 2.17  & 5.02  & 9.30 & 2.57 & 5.22 & 9.41 & 2.11 & 4.53 & 7.31 & 2.64 & 4.79 & 7.46 \\
$\langle C\rangle$      & 0.13  & 0.29  & 0.47 & 0.13 & 0.29 & 0.47 & 0.20 & 0.32 & 0.56 & 0.20 & 0.32 & 0.56 \\
Components              & 1     & 5     & 24   & 1   & 1   & 1   & 1   & 6   & 37   & 1   & 1   & 1 \\
$n/N$                   & --    & --    & --   & 81\% & 95\% & 100\% & -- & -- & -- & 50\% & 94\% & 100\% \\
$m/M$                   & --    & --    & --   & 97\% & 100\% & 100\% & -- & -- & -- & 65\% & 100\% & 100\% \\
\bottomrule
\end{tabular}

\vspace{2pt}
\begin{minipage}{0.98\linewidth}
\footnotesize
\emph{Notes:} $N$ = number of nodes; $M$ = number of edges; $\delta$ = edge density; $\langle k\rangle$ = mean degree;
$\langle C\rangle$ = mean clustering coefficient; \textit{Components} = number of connected components.
LCC = largest connected component; $n/N$ and $m/M$ are the fractions of nodes and edges contained in the LCC.
\end{minipage}
\end{table}

\subsection*{Health networks}
Table~\ref{tab:health_w1w3} reports results for the health layer. 
Median degree and density are lower than in friendship (Wave~1: \(\langle k\rangle=2.82\), \(\delta=0.035\); Wave~3: \(\langle k\rangle=2.45\), \(\delta=0.030\)), and networks are more fragmented (median of 11 and 13 components). 
The LCCs capture a median of \(86\%\) (Wave~1) and \(80\%\) (Wave~3) of nodes and \(99\%\) and \(96\%\) of ties, respectively. 
As expected, LCC densities are higher, and median mean degree in the LCC exceeds that of the full network by roughly \(16\%\) (Wave~1) and \(19\%\) (Wave~3), while median clustering remains essentially unchanged (within \(\sim\!2\%\)).

\begin{table}[H]
\centering
\footnotesize
\caption{Health layer (Wave~1 and Wave~3): characteristics of full networks and LCCs.}
\label{tab:health_w1w3}
\begin{tabular}{lccc ccc ccc ccc}
\toprule
& \multicolumn{6}{c}{\textbf{Wave 1}} & \multicolumn{6}{c}{\textbf{Wave 3}} \\
\cmidrule(lr){2-7}\cmidrule(lr){8-13}
& \multicolumn{3}{c}{Networks} & \multicolumn{3}{c}{LCC} & \multicolumn{3}{c}{Networks} & \multicolumn{3}{c}{LCC} \\
\cmidrule(lr){2-4}\cmidrule(lr){5-7}\cmidrule(lr){8-10}\cmidrule(lr){11-13}
\textbf{Metric} & Min & Med & Max & Min & Med & Max & Min & Med & Max & Min & Med & Max \\
\midrule
$N$                     & 16 & 80 & 370 & 10 & 64 & 284 & 16 & 80 & 370 & 7  & 59 & 307 \\
$M$                     & 18 & 115 & 536 & 14 & 113 & 504 & 13 & 98  & 456 & 6  & 92 & 429 \\
$\delta$                & 0.006 & 0.035 & 0.150 & 0.009 & 0.051 & 0.356 & 0.006 & 0.030 & 0.108 & 0.009 & 0.051 & 0.286 \\
$\langle k\rangle$      & 1.09  & 2.82  & 6.77 & 2.00 & 3.26 & 6.85 & 0.88 & 2.45 & 4.73 & 1.71 & 2.90 & 4.94 \\
$\langle C\rangle$      & 0.00  & 0.29  & 0.65 & 0.00 & 0.29 & 0.63 & 0.04 & 0.29 & 0.50 & 0.00 & 0.29 & 0.49 \\
Components              & 1     & 11    & 91   & 1   & 1   & 1   & 2   & 13   & 101  & 1   & 1   & 1 \\
$n/N$                   & --    & --    & --   & 29\% & 86\% & 100\% & -- & -- & -- & 19\% & 80\% & 97\% \\
$m/M$                   & --    & --    & --   & 42\% & 99\% & 100\% & -- & -- & -- & 32\% & 96\% & 100\% \\
\bottomrule
\end{tabular}

\vspace{2pt}
\begin{minipage}{0.98\linewidth}
\footnotesize
\emph{Notes:} $N$ = number of nodes; $M$ = number of edges; $\delta$ = edge density; $\langle k\rangle$ = mean degree;
$\langle C\rangle$ = mean clustering coefficient; \textit{Components} = number of connected components.
LCC = largest connected component; $n/N$ and $m/M$ are the fractions of nodes and edges contained in the LCC.
\end{minipage}
\end{table}

\subsection*{Financial networks}
As shown in Table~\ref{tab:financial_w1w3}, the financial layer exhibits low degree and modest clustering (median \(\langle k\rangle=2.41\) and \(2.26\); median \(\langle C\rangle=0.23\) and \(0.24\) in Waves~1 and~3, respectively), with substantial fragmentation (median 15 and 16 components). 
The LCC contains a median of \(79\%\) (Wave~1) and \(75\%\) (Wave~3) of nodes and \(96\%\) and \(95\%\) of ties. 
Relative to the full networks, LCC median degree is higher by about \(19\%\) (Wave~1) and \(28\%\) (Wave~3), while median clustering remains within a few percent.

\begin{table}[H]
\centering
\footnotesize
\caption{Financial layer (Wave~1 and Wave~3): characteristics of full networks and LCCs.}
\label{tab:financial_w1w3}
\begin{tabular}{lccc ccc ccc ccc}
\toprule
& \multicolumn{6}{c}{\textbf{Wave 1}} & \multicolumn{6}{c}{\textbf{Wave 3}} \\
\cmidrule(lr){2-7}\cmidrule(lr){8-13}
& \multicolumn{3}{c}{Networks} & \multicolumn{3}{c}{LCC} & \multicolumn{3}{c}{Networks} & \multicolumn{3}{c}{LCC} \\
\cmidrule(lr){2-4}\cmidrule(lr){5-7}\cmidrule(lr){8-10}\cmidrule(lr){11-13}
\textbf{Metric} & Min & Med & Max & Min & Med & Max & Min & Med & Max & Min & Med & Max \\
\midrule
$N$                     & 16 & 80 & 368 & 7  & 60 & 304 & 16 & 80 & 368 & 5  & 55 & 284 \\
$M$                     & 12 & 97 & 497 & 8  & 92 & 491 & 11 & 92 & 457 & 4  & 84 & 414 \\
$\delta$                & 0.005 & 0.030 & 0.117 & 0.011 & 0.057 & 0.381 & 0.006 & 0.029 & 0.109 & 0.010 & 0.056 & 0.500 \\
$\langle k\rangle$      & 1.00  & 2.41  & 5.90 & 1.88 & 2.88 & 6.20 & 0.74 & 2.26 & 4.73 & 1.60 & 2.88 & 5.04 \\
$\langle C\rangle$      & 0.00  & 0.23  & 0.44 & 0.00 & 0.23 & 0.45 & 0.00 & 0.24 & 0.61 & 0.00 & 0.23 & 0.58 \\
Components              & 2     & 15    & 96   & 1   & 1   & 1   & 2   & 16   & 97   & 1   & 1   & 1 \\
$n/N$                   & --    & --    & --   & 13\% & 79\% & 98\% & -- & -- & -- & 9\% & 75\% & 98\% \\
$m/M$                   & --    & --    & --   & 25\% & 96\% & 100\% & -- & -- & -- & 21\% & 95\% & 100\% \\
\bottomrule
\end{tabular}

\vspace{2pt}
\begin{minipage}{0.98\linewidth}
\footnotesize
\emph{Notes:} $N$ = number of nodes; $M$ = number of edges; $\delta$ = edge density; $\langle k\rangle$ = mean degree;
$\langle C\rangle$ = mean clustering coefficient; \textit{Components} = number of connected components.
LCC = largest connected component; $n/N$ and $m/M$ are the fractions of nodes and edges contained in the LCC.
\end{minipage}
\end{table}

\subsection*{Adversarial networks}
Finally, Table~\ref{tab:adversarial_w1w3} shows the adversarial layer, which is the sparsest and most fragmented. 
Median degree is below one (Wave~1: \(\langle k\rangle=0.88\); Wave~3: \(0.53\)), with extremely low densities (Wave~1 min \(\delta \approx 4.17\times 10^{-4}\)). 
Networks comprise many components (medians 45 and 59). 
The LCCs contain a relatively small share of nodes, a median of \(\mathbf{24\%}\) (Wave~1) and \(13\%\) (Wave~3), and of ties (\(67\%\) and \(50\%\)). 
Because the LCC is, by construction, connected, its median degree is substantially higher than the network median (roughly \(2.4\times\) in Wave~1 and \(3.6\times\) in Wave~3), while median clustering remains zero.

\begin{table}[H]
\centering
\footnotesize
\caption{Adversarial layer (Wave~1 and Wave~3): characteristics of full networks and LCCs.}
\label{tab:adversarial_w1w3}
\begin{tabular}{lccc ccc ccc ccc}
\toprule
& \multicolumn{6}{c}{\textbf{Wave 1}} & \multicolumn{6}{c}{\textbf{Wave 3}} \\
\cmidrule(lr){2-7}\cmidrule(lr){8-13}
& \multicolumn{3}{c}{Networks} & \multicolumn{3}{c}{LCC} & \multicolumn{3}{c}{Networks} & \multicolumn{3}{c}{LCC} \\
\cmidrule(lr){2-4}\cmidrule(lr){5-7}\cmidrule(lr){8-10}\cmidrule(lr){11-13}
\textbf{Metric} & Min & Med & Max & Min & Med & Max & Min & Med & Max & Min & Med & Max \\
\midrule
$N$                     & 16 & 77 & 365 & 2  & 21 & 166 & 16 & 77 & 365 & 2  & 10 & 87 \\
$M$                     & 2  & 36 & 226 & 1  & 22 & 220 & 1  & 21 & 119 & 1  & 10 & 99 \\
$\delta$                & $4.17\times 10^{-4}$ & 0.011 & 0.067 & 0.016 & 0.108 & 1.000 & $5.23\times 10^{-4}$ & 0.006 & 0.034 & 0.026 & 0.222 & 1.000 \\
$\langle k\rangle$      & 0.06 & 0.88 & 2.53 & 1.00 & 2.08 & 3.21 & 0.07 & 0.53 & 1.81 & 1.00 & 1.89 & 2.75 \\
$\langle C\rangle$      & 0.00 & 0.00 & 0.56 & 0.00 & 0.00 & 0.58 & 0.00 & 0.00 & 0.43 & 0.00 & 0.00 & 0.70 \\
Components              & 8    & 45   & 268  & 1   & 1   & 1   & 15  & 59   & 288  & 1   & 1   & 1 \\
$n/N$                   & --   & --   & --   & 2\%  & 24\% & 81\% & -- & -- & -- & 2\% & 13\% & 72\% \\
$m/M$                   & --   & --   & --   & 13\% & 67\% & 100\% & -- & -- & -- & 9\% & 50\% & 100\% \\
\bottomrule
\end{tabular}

\vspace{2pt}
\begin{minipage}{0.98\linewidth}
\footnotesize
\emph{Notes:} $N$ = number of nodes; $M$ = number of edges; $\delta$ = edge density; $\langle k\rangle$ = mean degree;
$\langle C\rangle$ = mean clustering coefficient; \textit{Components} = number of connected components.
LCC = largest connected component; $n/N$ and $m/M$ are the fractions of nodes and edges contained in the LCC.
\end{minipage}
\end{table}

\subsection*{Cross-layer comparison}
Across layers and waves, network cohesion declines monotonically from \emph{friendship} to \emph{health} to \emph{financial} to \emph{adversarial}. 
Median edge density is highest in friendship (Wave~1: \(\delta=0.062\); Wave~3: \(\delta=0.058\)) with substantial clustering (\(\langle C\rangle \approx 0.29/0.32\)) and mild fragmentation (median components \(=5/6\)); the LCC captures almost the entire network (median \(n/N=95.1\%/94.4\%\), \(m/M=100\%/100\%\)). 
Health and financial layers are sparser and more fragmented (health: \(\delta \approx 0.035/0.030\), components \(=11/13\); financial: \(\delta \approx 0.0295/0.0289\), components \(=15/16\)), but their LCCs still contain the large majority of nodes and nearly all ties (health: \(n/N=85.9\%/79.6\%\), \(m/M=98.8\%/96.2\%\); financial: \(n/N=79.4\%/75.0\%\), \(m/M=96.4\%/94.7\%\)). 
In \emph{adversarial} networks, by contrast, density and clustering are minimal (\(\delta \approx 0.011/0.006\); \(\langle C\rangle=0\)), fragmentation is extreme (components \(=45/59\)), and the LCC covers only a small fraction of nodes and ties (median \(n/N=23.8\%/12.6\%\), \(m/M=66.7\%/50.0\%\)). 
Consistently across layers, the LCC median mean degree exceeds the full-network median, slightly in friendship (\(+4\%\) to \(+6\%\)), modestly in health (\(+16\%\) to \(+19\%\)) and financial (\(+19\%\) to \(+28\%\)), and dramatically in adversarial (\(+136\%\) to \(+258\%\)), while median clustering changes little between full networks and LCCs.

\section{Empirical Results}

\subsection{Cooperative layers co-move but are not redundant}
\label{sec:coupling-vs-overlap}

\begin{figure}[H]
  \centering
  \includegraphics[scale=0.4]{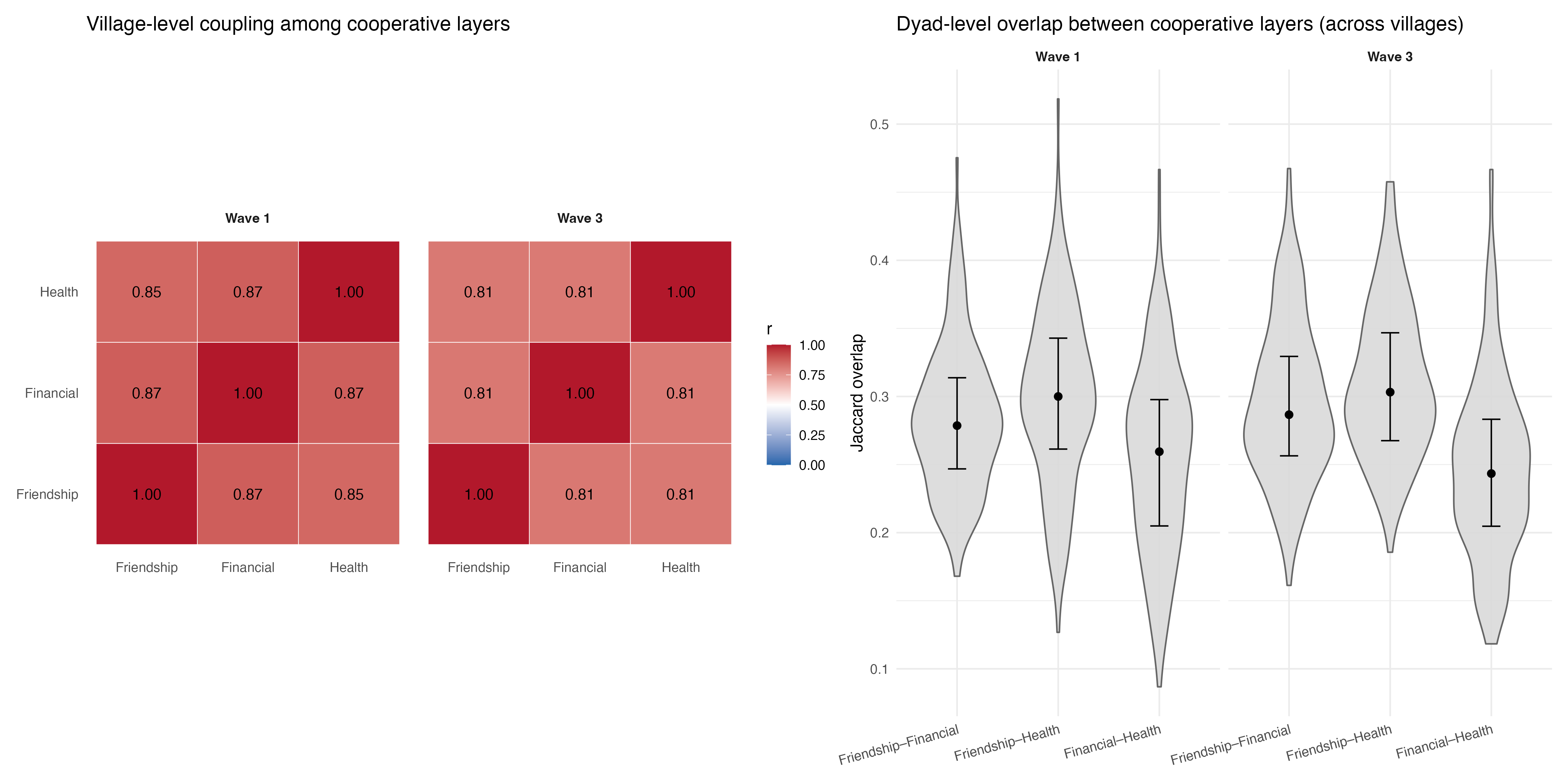}
  \caption{\textbf{Village‑level coupling and dyad‑level specialization of cooperative layers.}
  \emph{Left:} Correlations among village random effects from the multivariate degree model (Friendship, Health, Financial), faceted by wave (W1, W3). When random effects are unavailable, the panel uses per‑village mean degree as a proxy (results are nearly identical).
  \emph{Right:} Distribution of dyad‑level overlap (Jaccard index) between pairs of cooperative layers across villages, by wave. Violin shapes show the across‑village distribution; points mark medians.}
  \label{fig:coop_coupling_overlap}
\end{figure}

At the village level, the three cooperative layers co-move strongly. 
Using village random intercepts from the multivariate degree model in Eq.~\eqref{eq:degree_model}, we find high cross-layer correlations in both waves (typically $r \approx 0.86$--$0.92$), indicating that places with dense friendship networks also tend to have dense health and financial help networks. 
The left panel of Fig.~\ref{fig:coop_coupling_overlap} summarizes these correlations by wave.\\

At the dyad level, cooperative ties are functionally specialized rather than fully redundant. Jaccard overlaps between pairs of cooperative layers are modest in both waves, typically in the $0.20$–$0.40$ range across villages (right panel of Fig.~\ref{fig:coop_coupling_overlap}). 
Across villages, the median overlap is about $0.27$ for Friendship--Financial, $0.29$ for Friendship--Health,
and $0.26$ for Financial--Health (roughly $0.22$–$0.35$), with only small shifts from Wave~1 to Wave~3. Thus, even in socially rich places, richness does not arise from stacking multiple cooperative relations on the same pairs; instead, different cooperative functions are carried by partially overlapping but distinct dyads.

We estimate ego-level overlap between cooperative layers—Friendship, Financial (F--Fin), Friendship--Health (F--H), and Financial--Health (Fin--H), using beta mixed models with random intercepts for village and respondent and weights proportional to the pairwise union size. For each focal attribute, models include the other six covariates and wave fixed effects; the values below are adjusted marginal means averaged across waves. Overlaps are modest in absolute size (across all models, typically $0.10$--$0.28$), consistent with functional specialization, but who reuses the same partners across functions varies systematically:

\begin{itemize}
  \item \textbf{Gender:} Women show more overlap than men, especially where health is involved:
  F--H $=0.25$ (women) vs.\ $0.16$ (men; $\Delta=+0.09$),
  Fin--H $=0.16$ vs.\ $0.10$ ($\Delta=+0.06$),
  and a small edge in F--Fin ($0.21$ vs.\ $0.20$; $\Delta=+0.01$).
  \emph{Triple multiplexity:} not estimable in the GLMM (no stable contrasts returned).

  \item \textbf{Age:} Older respondents have more overlap across all pairs:
  F--H $=0.28$ (older) vs.\ $0.25$ (younger; $\Delta=+0.03$),
  F--Fin $=0.26$ vs.\ $0.24$ ($\Delta=+0.03$),
  Fin--H $=0.18$ vs.\ $0.16$ ($\Delta=+0.02$).
  \emph{Triple multiplexity:} $J_{\text{triple}}=0.15$ (older) vs.\ $0.13$ (younger; $\Delta=+0.02$);
  $\Pr(\text{any triple})=0.48$ vs.\ $0.33$ (i.e., $48\%$ vs.\ $33\%$; $\Delta\approx+15$ pp).

  \item \textbf{Education:} Education raises overlap when money is involved, but not for F--H:
  F--Fin $=0.27$ (educated) vs.\ $0.24$ (non-educated; $\Delta=+0.03$);
  Fin--H $=0.18$ vs.\ $0.17$ ($\Delta=+0.01$);
  F--H is slightly \emph{higher} among the non-educated ($0.26$) than the educated ($0.26$; $\Delta\approx-0.01$).
  \emph{Triple multiplexity:} $J_{\text{triple}}=0.14$ (educated) vs.\ $0.14$ (non; $\Delta\approx+0.01$);
  $\Pr(\text{any triple})=0.43$ vs.\ $0.38$ ($+5$ pp).

  \item \textbf{Income sufficiency:} Those reporting insufficient income show slightly more overlap when health is involved:
  F--H $=0.26$ (sufficient) vs.\ $0.28$ (insufficient; $\Delta\approx-0.02$ when coded sufficient$-$insufficient),
  Fin--H $=0.17$ vs.\ $0.18$ ($\Delta\approx-0.01$),
  and essentially no gap in F--Fin ($0.25$ vs.\ $0.25$).
  \emph{Triple multiplexity:} $J_{\text{triple}}=0.15$ (sufficient) vs.\ $0.14$ (insufficient; $\Delta\approx+0.00$);
  $\Pr(\text{any triple})=0.53$ vs.\ $0.53$ (no meaningful difference).

  \item \textbf{Marital status:} Singles overlap slightly more than partnered across all pairs:
  F--H $=0.20$ (single) vs.\ $0.20$ (partnered; $\Delta\approx+0.01$),
  F--Fin $=0.20$ vs.\ $0.19$ ($\Delta\approx+0.01$),
  Fin--H $=0.14$ vs.\ $0.13$ ($\Delta\approx+0.01$).
  \emph{Triple multiplexity:} not estimable in the triple-overlap models.

  \item \textbf{Indigenous identity:} Non-indigenous respondents have slightly higher pairwise overlaps:
  F--H $=0.16$ (non-indigenous) vs.\ $0.16$ (indigenous; $\Delta\approx+0.01$),
  F--Fin $=0.15$ vs.\ $0.14$ ($\Delta\approx+0.01$),
  Fin--H $=0.10$ vs.\ $0.09$ ($\Delta\approx+0.01$).
  \emph{Triple multiplexity:} $J_{\text{triple}}=0.12$ (non-indigenous) vs.\ $0.12$ (indigenous; $\Delta\approx+0.00$);
  $\Pr(\text{any triple})=0.37$ vs.\ $0.39$ (indigenous slightly higher, $+2$ pp).

  \item \textbf{Religion:} Protestants have the highest pairwise overlaps:
  F--H: Protestant $=0.23$ $\;>\;$ Catholic $=0.21$ $\approx$ None/Other $=0.21$;
  F--Fin: Protestant $=0.22$ $\;>\;$ Catholic $=0.20$ $\approx$ None/Other $=0.20$;
  Fin--H: Protestant $=0.15$ $\;>\;$ Catholic $=0.14$ $\;>\;$ None/Other $=0.13$.
  \emph{Triple multiplexity:} $J_{\text{triple}}$: Catholic $=0.14$ $\gtrsim$ Protestant $=0.13$ $>$ None/Other $=0.13$;
  $\Pr(\text{any triple})$: Catholic $=0.47$ $>$ Protestant $=0.44$ $>$ None/Other $=0.39$.
\end{itemize}

Overlaps are modest (pairwise $0.10$–$0.28$; triple $J_{\text{triple}}\approx0.12$–$0.15$; $\Pr(\text{any triple})\approx0.33$–$0.53$), so layers remain only partially redundant. Women, older adults, the educated (for money ties), and Protestants are the groups most likely to \emph{reuse} partners across functions; income, marital status, and indigenous identity show small or no differences.

\subsection{Degree models: who keeps ties and how layers co‑move}
\label{sec:degree_models}

Degree distributions by layer and wave are summarized in Table~\ref{tab:deg_by_layer}.

\begin{table}[H]
\centering
\caption{Degree distributions by layer and wave. All $n$ are respondents observed per wave.}
\label{tab:deg_by_layer}
\begin{tabular}{llrrrrr}
\toprule
Layer & Wave & $n$ & Mean & Variance & Zeros (n) & Zeros (\%) \\
\midrule
Health      & 1 & 20{,}232 & 2.84 & 10.30 & 3{,}691 & 18.2 \\
            & 3 & 20{,}232 & 2.40 & \;\;7.95 & 4{,}060 & 20.1 \\
Friendship  & 1 & 20{,}232 & 7.36 & 30.80 & 2{,}522 & 12.5 \\
            & 3 & 20{,}232 & 6.91 & 27.10 & 2{,}530 & 12.5 \\
Financial   & 1 & 20{,}232 & 2.73 & \;\;7.18 & 4{,}348 & 21.5 \\
            & 3 & 20{,}232 & 2.53 & \;\;6.58 & 4{,}664 & 23.1 \\
Adversarial & 1 & 20{,}232 & 0.90 & \;\;1.90 & 11{,}168 & 55.2 \\
            & 3 & 20{,}232 & 0.53 & \;\;1.04 & 14{,}190 & 70.2 \\
\bottomrule
\end{tabular}
\end{table}

We model respondent $i$'s degree $y_i^{(\ell)}$ in layer $\ell\in\{\text{Health},\ \text{Friendship},\ \text{Financial},\ \text{Adversarial}\}$ using a cross‑classified mixed specification with respondent, village, and wave random intercepts. For both cooperative and conflict layers we fit zero‑inflated NB2 with a log link. Coefficients are reported as incidence‑rate ratios (IRR). Formally,
\begin{equation}
\label{eq:degree_model}
\log \mathrm{E}\!\left[y_i^{(\ell)}\right]
= \beta_0
+ \beta_{\mathrm{fr}}\,\mathrm{fr}_i
+ \beta_{\mathrm{fin}}\,\mathrm{fin}_i
+ \beta_{\mathrm{adv}}\,\mathrm{adv}_i
+ \boldsymbol{\beta}^{\top}\mathbf{X}_i
+ u_{\text{village}[i]} + u_{\text{wave}[i]} + u_{\text{resp}[i]}\,,
\end{equation}
with a separate zero‑inflation equation for cooperative layers. Degrees are over‑dispersed and substantially zero‑inflated for cooperative layers (e.g., Friendship mean/var/zeros: W1 = 7.36/30.8/12.5\%; W3 = 6.91/27.1/12.5\%), while adversarial degrees are sparse (W1 mean/var/zeros = 0.90/1.90/55.2\%; W3 = 0.53/1.04/70.2\%), motivating zero inflated negative binomial model. Data details and full tables are presented in the SI.\\

Education is associated with more cooperative ties and fewer conflict ties: Health (IRR $\approx 1.010$ per year; $p<0.001$), Friendship (IRR $=1.009$, $p=1.19\times 10^{-9}$), Financial (IRR $=1.014$, $p=1.01\times 10^{-11}$), and Adversarial (IRR $=0.974$, $p=1.00\times 10^{-9}$). More schooling is linked to broader supportive networks and fewer antagonistic relations, possibly consistent with better communication/status and conflict‑management skills. The pattern that education expands cooperative networks and reduces conflict is consistent with other research linking schooling to larger and more diverse personal networks and civic engagement \citep{bokanyi2023anatomy,putnam2000bowling}. \\

Age increases Friendship (IRR $=1.004$, $p=1.20\times 10^{-60}$) and Financial degree (IRR $=1.002$, $p=1.75\times 10^{-13}$) and slightly lowers Adversarial degree (IRR $=0.996$, $p=1.94\times 10^{-6}$); Health shows a positive age gradient as well (IRR $\approx 1.008$ per year; $p<0.001$). Older adults sustain somewhat broader cooperative engagement but are marginally less involved in conflict, consistent with life‑cycle shifts toward stability and selective engagement. A modest positive age gradient in cooperative layers and reduced antagonistic ties accords with socioemotional selectivity theory and evidence of age‑structured network change in rural settings \citep{carstensen1992social,harling2020age}. Generally speaking, older adults prune low‑value ties and avoid conflict. \\

Men report more Financial ties (IRR $=1.234$, $p<2\times 10^{-16}$) but fewer Friendship ties (IRR $=0.968$, $p=2.04\times 10^{-6}$) and markedly fewer Adversarial ties (IRR $=0.624$, $p=3.99\times 10^{-66}$). Men’s instrumental exchange networks are larger, whereas friendship networks are slightly smaller; conflictual ties are concentrated in fewer named relations (or resolved/avoided), conditional on other covariates. Men’s larger financial out‑degree and slightly smaller friendship out‑degree are consistent with gendered role specialization in instrumental versus socio‑emotional ties; contextual evidence on women’s constrained networks in low‑income settings supports these differences \citep{andrew2020mothers,harling2020age}.\\

Marriage is positively associated with degree across layers: Friendship (IRR $=1.064$, $p<2\times 10^{-16}$), Financial (IRR $=1.083$, $p=6.95\times 10^{-16}$), and Adversarial (IRR $=1.076$, $p=2.43\times 10^{-4}$). Marriage appears to broaden outward engagement—spouses connect households to more partners across activities and also modestly raises named antagonistic ties, plausibly as a by‑product of wider interaction. The positive association of marriage with friendship and financial degree is compatible with household‑to‑household linking via spouses and in‑laws; small increases in antagonistic out‑degree may reflect greater exposure through wider interaction (see general social capital arguments in \citealp{putnam2000bowling}). \\

Income sufficiency reduces Friendship (IRR $=0.982$, $p=1.12\times 10^{-3}$) and Adversarial (IRR $=0.957$, $p=0.0106$) but increases Financial degree (IRR $=1.044$, $p=2.64\times 10^{-7}$). Economically secure respondents engage more in voluntary exchange networks, but name slightly fewer casual friends and conflicts, consistent with less necessity‑driven socializing and lower exposure to disputes. Higher financial degree alongside fewer friendship/conflict ties suggests a shift from necessity‑driven to elective engagement; related work documents socioeconomic gradients in local isolation and differing reliance on informal ties \citep{andrew2020mothers,bokanyi2023anatomy}.\\

Compared to the no religious villages, Protestants and Catholics report more Friendship ties (Protestant: IRR $=1.056$, $p=9.43\times 10^{-7}$; Catholic: IRR $=1.036$, $p=0.00155$) and more Financial ties (Protestant: IRR $=1.063$, $p=9.79\times 10^{-5}$; Catholic: IRR $=1.050$, $p=0.00242$); effects on Adversarial are small and not significant (Protestant: IRR $=0.994$, $p=0.862$; Catholic: IRR $=0.941$, $p=0.0719$). Membership in the two major denominations is associated with slightly denser cooperative networks, with no systematic differences in conflict. Modest cooperative advantages for Protestants/Catholics, with no clear conflict difference, fit the view that affiliation offers mild bonding capital without strongly re‑wiring everyday village ties; homophily may be present but limited in integrated communities \citep{lynch2022religiosity,mcpherson2001birds}. \\

Indigenous respondents report slightly more ties in each layer: Friendship (IRR $=1.026$, $p=0.0308$), Financial (IRR $=1.040$, $p=0.0207$), and Adversarial (IRR $=1.14$, $p=3.86\times 10^{-4}$). Conditional on other factors, indigenous villagers maintain modestly broader outward engagement and also name more antagonistic ties, consistent with greater overall interaction volume (more opportunities for both cooperation and friction). Slightly higher degrees across layers are consistent with greater interaction volume; in other contexts, minority status can yield strong bonding ties and weaker bridging ties \citep{woolcock1998social}.\\

Place correlates show limited average effects: access to road routes is not significant for Friendship (IRR $=1.025$, $p=0.339$), Financial (IRR $=1.033$, $p=0.128$), or Adversarial (IRR $=1.03$, $p=0.540$); village size is weakly negative for Friendship (IRR $=0.999$, $p=2.08\times 10^{-11}$) and near null for Financial and Adversarial (both $p>0.20$). Physical access alone does not reconfigure micro‑ties, while larger populations slightly dilute average friendship degree via segmentation. Net effects are context‑dependent. Null access‑to‑routes effects and a small negative village‑size association with friendship align with the idea that larger populations segment networks, diluting average degree \citep{dunbar1992neocortex,shakya2019village}. Physical access alone may be insufficient to alter micro‑tie formation.

\paragraph{How layers co‑move.}
Cooperative layers co‑move strongly. For the Friendship outcome, Financial degree has a large positive association (IRR $=1.10$, $p<2\times 10^{-16}$) and Health degree is also positive (IRR $=1.046$, $p<2\times 10^{-16}$). For the Financial outcome, Friendship (IRR $=1.073$, $p<2\times 10^{-16}$) and Health (IRR $=1.039$, $p<2\times 10^{-16}$) raise expected out‑degree. Health degree rises only slightly with cooperative degrees, Friendship (IRR $=1.068$, $p<2\times 10^{-16}$), Financial (IRR $=1.077$, $p<2\times 10^{-16}$), Adversarial (IRR $=1.034$, $p<2\times 10^{-16}$), indicating stronger co‑movement with conflict ties than among cooperative layers. Adversarial degree rises only slightly with cooperative degrees, Friendship (IRR $=1.021$, $p<2\times 10^{-16}$), Financial (IRR $=1.043$, $p<2\times 10^{-16}$), Health (IRR $=1.019$, $p=2.08\times 10^{-10}$), indicating far weaker co‑movement with conflict ties than among cooperative layers. \\

Between‑village heterogeneity (non‑zero random intercept variance) suggests contextual differences in overall connectedness. In short, socially active individuals tend to be active across cooperative domains (multiplexity), while antagonistic ties scale much less with cooperation. In the literature, strong co‑movement among cooperative layers and weak coupling with conflict is a hallmark of multiplex social organization: general social “hubs” accumulate ties across activities, whereas antagonism scales less \citep{gluckman1973judicial,ghasemian2024structure}.

\paragraph{Within vs.\ between and heterogeneity.}
A Mundlak decomposition distinguishes within‑ from between‑person gradients (e.g., for Health, the between‑person age effect is positive and precise; village size shows small negative within/between effects). Allowing village‑varying cross‑layer slopes reveals heterogeneity in how general connectedness translates into Health support: the SD of the Friendship to Health slope is $\hat\sigma=0.071$ on the log scale, i.e., a $\pm 7\%$ band around the mean effect. 

\paragraph{Wave differences.}
Wave~3 shows a mild contraction in cooperative degree (and stable zeros in Friendship) alongside a marked reduction and higher zero mass in Adversarial.\\

\begin{table}[H]
\centering
\caption{Core fixed‑effect IRRs from the degree models (count parts). Entries are shown only when $p<0.05$ in the corresponding layer‑specific model; otherwise the cell is left as “—”. Values are incidence‑rate ratios (IRR).}
\label{tab:deg-irrs-core}
\small
\begin{tabular}{lcccc}
\toprule
\textbf{Predictor} & \textbf{Health} & \textbf{Friendship} & \textbf{Financial} & \textbf{Adversarial} \\
\midrule
Friendship degree ($fr_c$)       & 1.058 & —     & 1.073 & 1.022 \\
Financial degree ($fin_c$)       & 1.077 & 1.101 & —     & 1.044 \\
Health degree ($h_c$)            & —     & 1.046 & 1.039 & 1.019 \\
Adversarial degree ($adv_c$)     & 1.034 & 1.022 & 1.040 & —     \\
\addlinespace
Education (per year)             & 1.014 & 1.009 & 1.014 & 0.974 \\
Age (per year)                   & 1.009 & 1.004 & 1.002 & 0.999 \\
Male                              & 0.687 & 0.968 & 1.234 & 0.629 \\
Married                           & 1.061 & 1.064 & 1.083 & —     \\
Income sufficient                 & 0.979 & 0.982 & 1.044 & 0.959 \\
Protestant (Rel.~level 1)         & 1.082 & 1.056 & 1.063 & —     \\
Catholic (Rel.~level 2)           & 1.051 & 1.037 & 1.050 & —     \\
Indigenous                        & 1.107 & 1.026 & 1.040 & 1.153 \\
Access to routes (centered)       & —     & —     & —     & —     \\
Village size (centered)           & —     & 0.999 & —     & —     \\
\bottomrule
\end{tabular}
\end{table}

\noindent Overall: (1) Cooperative layers co‑move strongly at both person and place levels (5–9\% per added tie; village $r{=}0.86$–$0.92$), yet dyadic overlap is low. 
(2) Education and age push cooperative and adversarial degrees in opposite directions; gender roles are specialized (men: financial; women: friendship/health; men report fewer adversarial ties).  
(3) Cross‑layer couplings vary across villages (±7\% around the mean for Friendship$\rightarrow$Health), highlighting scope for village‑level meta‑analysis and targeting. For a compact summary of the main covariate effects, see Table~\ref{tab:deg-irrs-core}.

\subsection{Dyad-level assortativity and degree-aware permutations}

We provide an in-depth analysis of factors associated with network connections in 176 rural villages in the Copan Department of Honduras, focusing on health, friendship, financial, and adversarial relationships across two years. Using mixed-effects zero inflated negative binomial regression models and assortativity measures, we examined how individual attributes and network characteristics associated with the formation of various social ties. We find that gender and religion are the most significant factors driving both within-community and inter-community assortativity. Specifically, strong gender homophily is evident across all network types, with female-female relationships predominating in health and friendship networks and male-male ties being more prevalent in financial networks. Additionally, shared religious affiliations significantly contribute to community cohesion and the formation of inter-community connections, highlighting the fundamental role of cultural and identity-based factors in shaping social structures.\\

Friendship and financial connections emerge as critical predictors across all types of relationships, underscoring their centrality in facilitating broader social interactions. Individuals with extensive friendship and financial networks are more likely to engage in multiple forms of social relationships, demonstrating the interconnected nature of social and economic ties in fostering community resilience and support systems. In contrast, attributes such as age, education, and income sufficiency exhibit weaker assortativity, particularly in inter-community ties. While these factors contribute to internal community cohesion, their limited role in bridging different communities suggests the presence of socio-economic stratifications or cultural barriers that hinder broader social integration. Community-level assortativity further emphasizes the importance of education, age, and religion in structuring communities, with increasing mutual information values over time indicating a growing alignment of community structures with socio-economic factors.\\

Permutation-based assortativity results are summarized in Table~\ref{tab:assort_all_layers}.

\begin{table}[H]
\centering
\caption{Assortativity across layers and tolerances. Cells show the percentage of villages with $p<0.05$ at tolerances $\pm5\%/\pm10\%/\pm20\%$ on group–specific mean degrees. For non‑gender attributes (Religion), permutations are stratified by gender.}
\label{tab:assort_all_layers}
\small
\begin{tabular}{lccccc}
\toprule
& \multicolumn{3}{c}{\textbf{Gender}} & \multicolumn{2}{c}{\textbf{Religion}} \\
\cmidrule(lr){2-4}\cmidrule(lr){5-6}
\textbf{Layer (Wave 1)} & M--M & F--F & M--F (dissort.) & P--P & C--C \\
\midrule
Friendship  & 97.1 / 96.6 / 92.0 & 96.0 / 93.1 / 74.3 & 0.0 / 0.0 / 0.0  & 67.4 / 64.0 / 60.0 & 60.0 / 54.9 / 48.6 \\
Health      & 97.1 / 36.6 / 16.6 & 96.0 / 38.9 / 50.9 & 0.6 / 0.0 / 0.0  & 60.0 / 58.9 / 54.9 & 53.7 / 46.9 / 36.6 \\
Financial   & 97.1 / 77.7 / 74.9 & 96.0 / 63.4 / 25.7 & 0.0 / 0.0 / 0.0  & 56.6 / 53.1 / 46.3 & 54.3 / 52.0 / 43.4 \\
Adversarial & 97.1 / 59.4 / 60.6 & 96.0 / 57.7 / 59.4 & 0.0 / 0.0 / 0.0  & 60.0 / 58.9 / 54.9 & 53.7 / 46.9 / 36.6 \\
\bottomrule
\end{tabular}
\end{table}




\subsection{Communities align most with friendship and are weakly explained by single traits}

We aggregate friendship, health, and financial ties by union to obtain denser undirected graphs suited for meso‑scale analysis. Table~\ref{tab:agg_w1w3} reports Min/Median/Max metric across villages for Wave~1 and Wave~3 (village~156 excluded). In Wave~1, aggregates are slightly denser (median $\delta{=}0.083$) with higher median mean degree ($\langle k\rangle{=}6.58$) and strong clustering ($\langle C\rangle{=}0.349$), while fragmentation is limited (median 4 components). Wave~3 shows a similar pattern (median $\delta{=}0.075$, $\langle k\rangle{=}6.02$, $\langle C\rangle{=}0.374$; median 5 components). In both waves the largest connected component (LCC) contains nearly the entire network (median $n/N{=}96.3\%$ in W1 and $95.7\%$ in W3; median $m/M{=}100\%$ in both) and is even denser (W1: $\delta_{\text{LCC}}{=}0.091$, $\langle k\rangle_{\text{LCC}}{=}6.86$; W3: $\delta_{\text{LCC}}{=}0.083$, $\langle k\rangle_{\text{LCC}}{=}6.32$), confirming that aggregation yields favorable conditions for community detection.

\begin{table}[H]
\centering
\footnotesize
\caption{Aggregated layer (Wave~1 and Wave~3; friendship $\cup$ health $\cup$ financial): characteristics of full networks and \textbf{largest connected components (LCCs)} (Min/Median/Max across villages; village 156 excluded).}
\label{tab:agg_w1w3}
\begin{tabular}{lccc ccc ccc ccc}
\toprule
& \multicolumn{6}{c}{\textbf{Wave 1 (Aggregated)}} & \multicolumn{6}{c}{\textbf{Wave 3 (Aggregated)}} \\
\cmidrule(lr){2-7}\cmidrule(lr){8-13}
& \multicolumn{3}{c}{Networks} & \multicolumn{3}{c}{LCC} & \multicolumn{3}{c}{Networks} & \multicolumn{3}{c}{LCC} \\
\cmidrule(lr){2-4}\cmidrule(lr){5-7}\cmidrule(lr){8-10}\cmidrule(lr){11-13}
\textbf{Metric} & Min & Med & Max & Min & Med & Max & Min & Med & Max & Min & Med & Max \\
\midrule
$N$ & 16 & 80 & 370 & 15 & 76 & 362 & 16 & 80 & 370 & 15 & 75 & 366 \\
$M$ & 29 & 274 & 1352 & 29 & 274 & 1351 & 41 & 237 & 1013 & 41 & 237 & 1013 \\
$\delta$ & 0.014 & 0.083 & 0.262 & 0.016 & 0.091 & 0.276 & 0.014 & 0.075 & 0.342 & 0.015 & 0.083 & 0.390 \\
$\langle k\rangle$ & 3.09 & 6.58 & 13.13 & 3.40 & 6.86 & 13.29 & 2.44 & 6.02 & 10.41 & 2.97 & 6.32 & 10.63 \\
$\langle C\rangle$ & 0.230 & 0.349 & 0.520 & 0.230 & 0.349 & 0.520 & 0.263 & 0.374 & 0.590 & 0.263 & 0.374 & 0.590 \\
Components & 1 & 4 & 21 & 1 & 1 & 1 & 1 & 5 & 22 & 1 & 1 & 1 \\
$n/N$ & -- & -- & -- & 84.3\% & 96.3\% & 100\% & -- & -- & -- & 77.1\% & 95.7\% & 100\% \\
$m/M$ & -- & -- & -- & 99.2\% & 100\% & 100\% & -- & -- & -- & 87.0\% & 100\% & 100\% \\
\bottomrule
\end{tabular}

\vspace{2pt}
\begin{minipage}{0.98\linewidth}
\footnotesize
\emph{Notes:} $N$ = number of nodes; $M$ = number of edges; $\delta$ = edge density; $\langle k\rangle$ = mean degree;
$\langle C\rangle$ = mean clustering coefficient; \textit{Components} = number of connected components.
LCC = largest connected component; $n/N$ and $m/M$ are the fractions of nodes and edges contained in the LCC.
\end{minipage}
\end{table}

\noindent We extract each village’s LCC, detect communities with Louvain ($\gamma{=}1$), and mask communities with $<4$ nodes, based on modularity. For a single (aggregated) layer with adjacency $A$, degrees $k$, $m$ edges, and partition $g$, modularity is
\begin{equation}
\label{eq:modularity}
Q \;=\; \frac{1}{2m}\sum_{i,j}\Big(A_{ij}-\gamma\,\frac{k_i k_j}{2m}\Big)\,\mathbf{1}\{g_i=g_j\},
\end{equation}
with $\gamma{=}1$ throughout. Applying Louvain ($\gamma\!=\!1$) on the LCC with a minimum community size of four yields a consistent meso-scale decomposition across villages and waves. Villages contain on average about six communities (W1: mean $6.12$, SD $1.84$, range $[2,12]$; W3: mean $6.26$, SD $1.92$, range $[3,15]$), indicating nontrivial segmentation beyond a core–periphery split. The largest community in a village typically holds about one quarter of covered nodes (W1: mean size $20.7$; mean share $0.251$; range $[7,60]$; W3: mean size $20.4$; mean share $0.247$; range $[5,55]$), so no single group dominates. Pooling communities over all villages, typical community size is $\sim$15 nodes with moderate dispersion (W1: mean $15.1$, SD $7.83$, range $[4,60]$; W3: mean $14.8$, SD $7.83$, range $[4,55]$). These structure-only results confirm that the aggregated networks admit stable, meso-scale community structure in both waves, setting the stage for the trait analyses below.

\begin{table}[H]
\centering
\footnotesize
\caption{Structure-only community statistics (Louvain on aggregated layer; LCC; minimum community size $=4$).}
\label{tab:comm_struct_only}
\begin{minipage}{0.35\textwidth}
\centering
\textbf{(A) \# communities per village}\\[2.5pt]
\begin{tabular}{lrrrrr}
\toprule
Wave & Vill. & Mean & SD & Min & Max \\
\midrule
1 & 176 & 6.12 & 1.84 & 2 & 12 \\
3 & 176 & 6.26 & 1.92 & 3 & 15 \\
\bottomrule
\end{tabular}
\end{minipage}\hfill
\begin{minipage}{0.35\textwidth}
\centering
\textbf{(B) Largest community per village}\\[2.5pt]
\begin{tabular}{lrrrr}
\toprule
Wave & Mean size & SD & Min & Max \\
\midrule
1 & 20.7 & 8.84 & 7 & 60 \\
3 & 20.4 & 9.24 & 5 & 55 \\
\midrule
\multicolumn{5}{l}{\small Mean share: W1 $=0.251$, W3 $=0.247$}\\
\bottomrule
\end{tabular}
\end{minipage}\hfill
\begin{minipage}{0.35\textwidth}
\centering
\textbf{(C) All communities pooled}\\[2.5pt]
\begin{tabular}{lrrrrr}
\toprule
Wave & Comms & Mean & SD & Min & Max \\
\midrule
1 & 1070 & 15.1 & 7.83 & 4 & 60 \\
3 & 1080 & 14.8 & 7.83 & 4 & 55 \\
\bottomrule
\end{tabular}
\end{minipage}
\end{table}

Structure-only community statistics are reported in Table~\ref{tab:comm_struct_only}.\\

\begin{figure}[H]
   \begin{minipage}{0.45\textwidth}
    \centering
    \includegraphics[width=\linewidth]{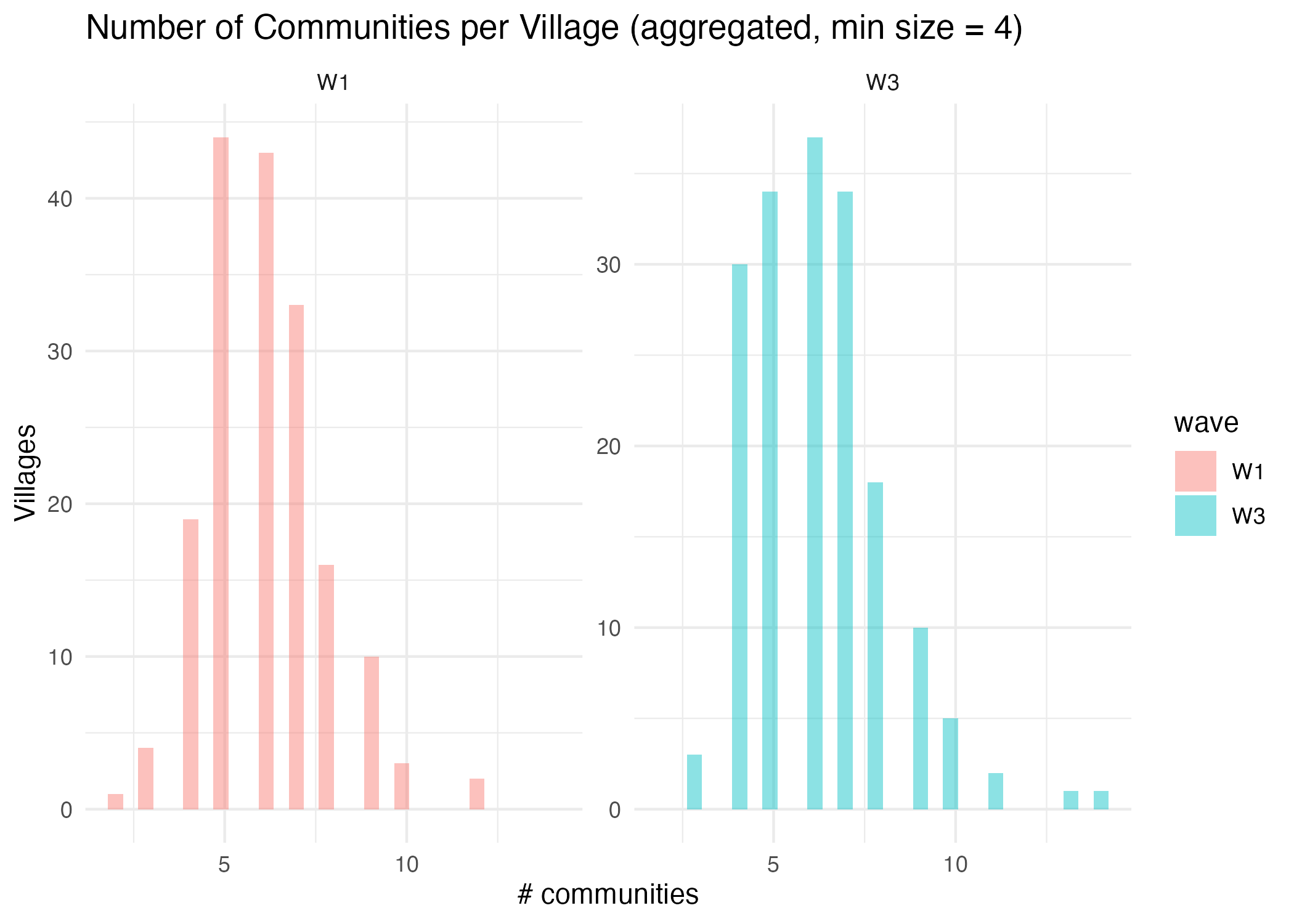}
  \end{minipage}
  \hspace{0.05\textwidth}
  \begin{minipage}{0.45\textwidth}
    \centering
    \includegraphics[width=\linewidth]{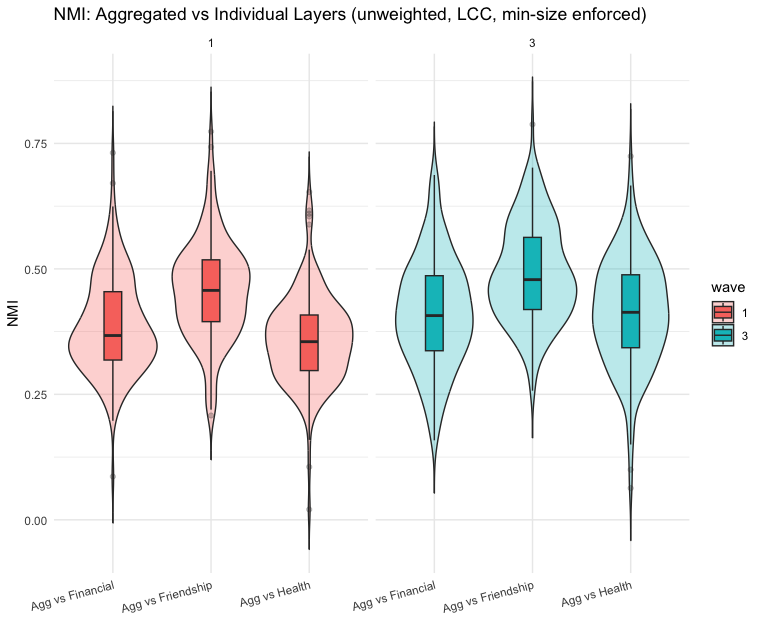}
  \end{minipage}

    \vspace{0.5cm}

  \begin{minipage}{0.45\textwidth}
    \centering
    \includegraphics[width=\linewidth]{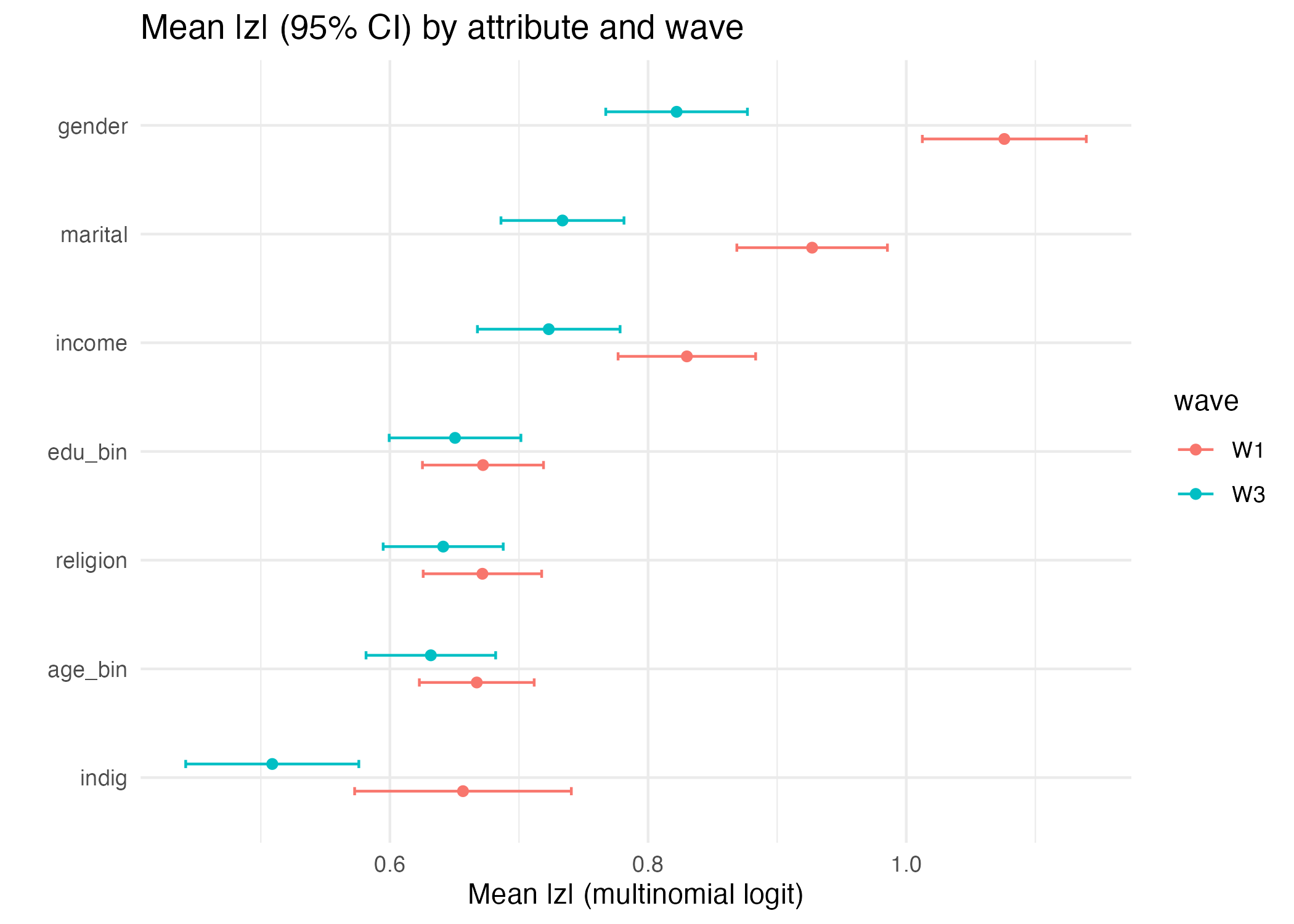}
  \end{minipage}%
  \hspace{0.05\textwidth}
  \begin{minipage}{0.45\textwidth}
    \centering
    \includegraphics[width=\linewidth]{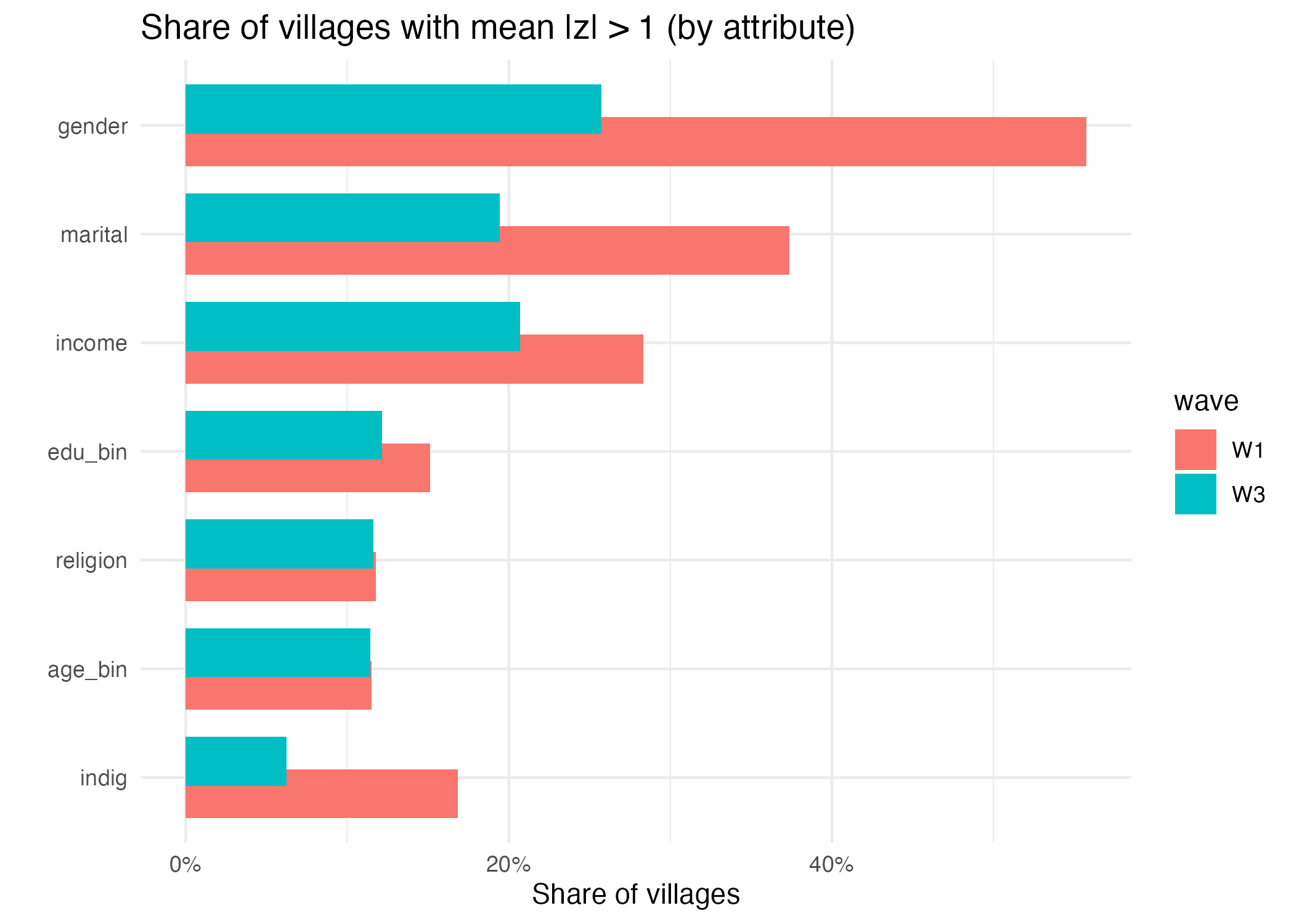}
  \end{minipage}

\caption{Upper Left: Number of communities per village (aggregated, min size $=4$).
           Upper Right: Alignment (NMI) between communities and single attributes.
           Lower Left: Mean $|z|$ (95\% CI) from within‑village multinomial logit.
           Lower Right: Share of villages with mean $|z|{>}1$ by attribute.}
\label{fig:nmi_top}
\end{figure}

\noindent Next, trait–community association is quantified in two complementary ways. 
(i) \emph{Prediction (multinomial logit).} Within each village we fit an $\ell_2$‑regularized multinomial model
\[
\text{Community} \sim \text{Gender} + \text{Religion} + \text{Marital Status} + \text{Indigenous Status} + \text{Income Sufficiency} + \text{Age} + \text{Education},
\]
and summarize each attribute’s contribution by the village‑wise mean absolute $z$ across its coefficients (``mean $|z|$''); we also report pseudo‑$R^2$ versus an intercept‑only model and
(ii) \emph{Alignment/permutation.} For each attribute we compute the modularity of the attribute partition on the aggregated graph ($Q_{\text{attr}}$) and assess significance with degree‑aware label permutations that preserve group‑specific mean degrees under $\pm 5\%$, $\pm 10\%$, and $\pm 20\%$ tolerances. We also visualize the normalized mutual information between Louvain communities and single attributes (NMI) to show direct label alignment (upper‑right panel of Fig.~\ref{fig:nmi_top}).\\

\noindent Across villages and waves, \emph{gender} is the strongest single predictor (highest median mean $|z|$), followed by \emph{marital status}, \emph{education}, and \emph{age}; \emph{income} and \emph{religion} are weaker on average. Pseudo‑$R^2$ is modest and stable across waves (means $\approx 0.25$). See Table~\ref{tab:comm_logit_summary} for across‑village summaries, and the lower panels of Fig.~\ref{fig:nmi_top} for mean $|z|$ (95\% CIs) and the share of villages with mean $|z|{>}1$.\\

\noindent Information‑theoretic alignment between community labels and any single attribute is small in median terms, as shown by the NMI panel (Fig.~\ref{fig:nmi_top}, upper‑right). Median NMI is near zero for most attributes, reflecting many villages with degenerate comparisons (e.g., one community after size masking or a single observed attribute level); a small but non‑zero median is visible for education. In contrast, $Q_{\text{attr}}$ is frequently positive and statistically detectable under degree‑aware permutations for \emph{gender} and \emph{religion}, and to a lesser extent for \emph{marital status} (Table~\ref{tab:comm_perm_summary}). \\

\noindent Overall, these results indicate that while several traits are assortative on edges (captured by $Q_{\text{attr}}$), they explain only limited variance in full community assignments (low NMI), consistent with communities reflecting multiple overlapping attributes rather than a single dominant trait.

\medskip

\begin{table}[H]
\centering
\caption{Predictive strength for community membership (mean $|z|$) and pseudo-$R^{2}$.}
\label{tab:comm_logit_summary}
\footnotesize
\begin{tabular}{llrrrr}
\toprule
Wave & Attribute & Villages & Median $|z|$ & Mean $|z|$ & Mean $R^2$ \\
\midrule
W1 & Gender & 176 & 1.030 & 1.070 & 0.252 \\
W1 & Marital & 176 & 0.850 & 0.930 & 0.252 \\
W1 & Income & 174 & 0.788 & 0.826 & 0.252 \\
W1 & Religion & 154 & 0.639 & 0.664 & 0.258 \\
W1 & Education (bin) & 172 & 0.612 & 0.673 & 0.251 \\
W1 & Age (bin) & 176 & 0.592 & 0.662 & 0.252 \\
W1 & Indigenous & 95 & 0.503 & 0.653 & 0.251 \\
W3 & Gender & 176 & 0.780 & 0.820 & 0.250 \\
W3 & Marital & 176 & 0.697 & 0.734 & 0.250 \\
W3 & Income & 174 & 0.629 & 0.724 & 0.250 \\
W3 & Religion & 155 & 0.588 & 0.641 & 0.253 \\
W3 & Education (bin) & 173 & 0.581 & 0.651 & 0.249 \\
W3 & Age (bin) & 176 & 0.566 & 0.629 & 0.250 \\
W3 & Indigenous & 96 & 0.448 & 0.510 & 0.253 \\
\bottomrule
\end{tabular}
\end{table}

\begin{table}[H]
\centering
\caption{Permutation summary. $Q_{\text{attr}}$ is the modularity of the attribute partition on the aggregated graph; shares are fractions of villages with $p<0.05$ under degree‑aware label permutations (10\% tolerance). NMI columns are omitted to avoid degenerate cases; NMI distributions are shown in Fig.~\ref{fig:nmi_top} (upper‑right).}
\label{tab:comm_perm_summary}
\footnotesize
\begin{tabular}{llrrr}
\toprule
Wave & Attribute & Villages & Median $Q_{\text{attr}}$ & Share $p^{Q}_{0.05}$ (10\%) \\
\midrule
W1 & Gender & 176 & 0.085 & 0.93 \\
W1 & Marital & 176 & 0.040 & 0.63 \\
W1 & Income & 176 & 0.004 & 0.15 \\
W1 & Religion & 176 & 0.043 & 0.74 \\
W1 & Education (bin) & 176 & 0.026 & 0.42 \\
W1 & Age (bin) & 176 & -0.011 & 0.28 \\
W1 & Indigenous & 176 & 0.000 & 0.40 \\
W3 & Gender & 176 & 0.073 & 0.91 \\
W3 & Marital & 176 & 0.025 & 0.40 \\
W3 & Income & 176 & 0.008 & 0.23 \\
W3 & Religion & 176 & 0.051 & 0.68 \\
W3 & Education (bin) & 176 & 0.020 & 0.33 \\
W3 & Age (bin) & 176 & -0.013 & 0.26 \\
W3 & Indigenous & 176 & 0.001 & 0.44 \\
\bottomrule
\end{tabular}
\end{table}


\section{Conclusion}

We provide an in-depth analysis of factors associated with network connections in 176 rural villages in the isolated Copan Department of Honduras, focusing on health, friendship, financial, and adversarial relationships across two years. Using mixed-effects zero inflated negative binomial regression models and assortativity measures, we examined how individual attributes and network characteristics are associated with the formation of various social ties. We find that gender and religion are the most significant factors driving both within network community assortativity and between network community (inter network community) assortativity. Specifically, strong gender homophily is evident across all network types, with female-female relationships predominating in health and friendship networks and male-male ties being more prevalent in financial networks. Additionally, shared religious affiliations significantly contribute to community cohesion and the formation of inter-community connections, highlighting the fundamental role of cultural and identity-based factors in shaping social structures.\\

\noindent Friendship and financial connections emerge as critical predictors across all types of relationships, underscoring their centrality in facilitating broader social interactions. Individuals with extensive friendship and financial networks are more likely to engage in multiple forms of social relationships, demonstrating the interconnected nature of social and economic ties in fostering community resilience and support systems. In contrast, attributes such as age, education, and income sufficiency exhibit weaker assortativity, particularly in inter network community ties. While these factors contribute to internal community cohesion, their limited role in bridging different communities suggests the presence of socio-economic stratifications or cultural barriers that hinder broader social integration. Community-level assortativity further emphasizes the importance of education, age, and religion in structuring communities, with increasing mutual information values over time indicating a growing alignment of community structures with socio-economic factors.\\

\noindent These underlying social patterns have implications for designing targeted interventions aimed at improving outcomes. By identifying and leveraging key individuals with extensive friendship and financial networks, health initiatives can more effectively disseminate information and resources, hereby enhancing community-wide health outcomes \citep{kim2015social,airoldi2024induction}. The observed strengthening of community cohesion based on education and income sufficiency over time suggests that socio-economic advancements play a crucial role in shaping resilient communities. Future research should explore the underlying mechanisms driving these assortativity patterns and investigate how ongoing socio-economic and cultural shifts influence network dynamics. Understanding these factors will provide deeper insights into sustaining community resilience and optimizing health interventions in similar rural settings.

\bibliographystyle{apalike}
\bibliography{references_1}

@article{montes2018connected,
  title={Connected but segregated: social networks in rural villages},
  author={Montes, Felipe and Jimenez, Roberto C and Onnela, Jukka-Pekka},
  journal={Journal of Complex Networks},
  volume={6},
  number={5},
  pages={693--705},
  year={2018},
  publisher={Oxford University Press}
}

@article{banerjee2013diffusion,
  title={The diffusion of microfinance},
  author={Banerjee, Abhijit and Chandrasekhar, Arun G and Duflo, Esther and Jackson, Matthew O},
  journal={Science},
  volume={341},
  number={6144},
  pages={1236498},
  year={2013},
  publisher={American Association for the Advancement of Science}
}

@article{lungeanu2021using,
  title={Using Trellis software to enhance high-quality large-scale network data collection in the field},
  author={Lungeanu, Alina and McKnight, Mark and Negron, Rennie and Munar, Wolfgang and Christakis, Nicholas A and Contractor, Noshir S},
  journal={Social Networks},
  volume={66},
  pages={171--184},
  year={2021},
  publisher={Elsevier}
}

@article{centola2011experimental,
  title={An experimental study of homophily in the adoption of health behavior},
  author={Centola, Damon},
  journal={Science},
  volume={334},
  number={6060},
  pages={1269--1272},
  year={2011},
  publisher={American Association for the Advancement of Science}
}

@article{kim2015social,
  title={Social network targeting to maximise population behaviour change: a cluster randomised controlled trial},
  author={Kim, David A and Hwong, Alison R and Stafford, Derek and Hughes, D Alex and O'Malley, A James and Fowler, James H and Christakis, Nicholas A},
  journal={The Lancet},
  volume={386},
  number={9989},
  pages={145--153},
  year={2015},
  publisher={Elsevier}
}

@article{hunter2015hidden,
  title={“Hidden” social networks in behavior change interventions},
  author={Hunter, Ruth F and McAneney, Helen and Davis, Michael and Tully, Mark A and Valente, Thomas W and Kee, Frank},
  journal={American journal of public health},
  volume={105},
  number={3},
  pages={513--516},
  year={2015},
  publisher={American Public Health Association}
}

@article{aral2013engineering,
  title={Engineering social contagions: Optimal network seeding in the presence of homophily},
  author={Aral, Sinan and Muchnik, Lev and Sundararajan, Arun},
  journal={Network Science},
  volume={1},
  number={2},
  pages={125--153},
  year={2013},
  publisher={Cambridge University Press}
}

@article{girvan2002community,
  title={Community structure in social and biological networks},
  author={Girvan, Michelle and Newman, Mark EJ},
  journal={Proceedings of the national academy of sciences},
  volume={99},
  number={12},
  pages={7821--7826},
  year={2002},
  publisher={The National Academy of Sciences}
}

@article{fortunato2010community,
  title={Community detection in graphs},
  author={Fortunato, Santo},
  journal={Physics reports},
  volume={486},
  number={3-5},
  pages={75--174},
  year={2010},
  publisher={Elsevier}
}

@article{blondel2008fast,
  title={Fast unfolding of communities in large networks},
  author={Blondel, Vincent D and Guillaume, Jean-Loup and Lambiotte, Renaud and Lefebvre, Etienne},
  journal={Journal of statistical mechanics: theory and experiment},
  volume={2008},
  number={10},
  pages={P10008},
  year={2008},
  publisher={IOP Publishing}
}

@article{newman2003mixing,
  title={Mixing patterns in networks},
  author={Newman, Mark EJ},
  journal={Physical review E},
  volume={67},
  number={2},
  pages={026126},
  year={2003},
  publisher={APS}
}

@article{noldus2015assortativity,
  title={Assortativity in complex networks},
  author={Noldus, Rogier and Van Mieghem, Piet},
  journal={Journal of complex networks},
  volume={3},
  number={4},
  pages={507--542},
  year={2015},
  publisher={Oxford University Press}
}

@article{henry2011emergence,
  title={Emergence of segregation in evolving social networks},
  author={Henry, Adam Douglas and Pra{\l}at, Pawe{\l} and Zhang, Cun-Quan},
  journal={Proceedings of the National Academy of Sciences},
  volume={108},
  number={21},
  pages={8605--8610},
  year={2011},
  publisher={National Academy of Sciences}
}

@article{kumpula2007emergence,
  title={Emergence of communities in weighted networks},
  author={Kumpula, Jussi M and Onnela, Jukka-Pekka and Saram{\"a}ki, Jari and Kaski, Kimmo and Kert{\'e}sz, J{\'a}nos},
  journal={Physical review letters},
  volume={99},
  number={22},
  pages={228701},
  year={2007},
  publisher={APS}
}

@article{porter2009communities,
  title={Communities in networks},
  author={Porter, Mason Alexander and Onnela, Jukka-Pekka and Mucha, Peter J and others},
  year={2009},
  publisher={SSRN}
}

@article{munshi2006traditional,
  title={Traditional institutions meet the modern world: Caste, gender, and schooling choice in a globalizing economy},
  author={Munshi, Kaivan and Rosenzweig, Mark},
  journal={American Economic Review},
  volume={96},
  number={4},
  pages={1225--1252},
  year={2006},
  publisher={American Economic Association}
}

@article{sanyal2015group,
  title={Group-based microcredit \& emergent inequality in social capital: Why socio-religious composition matters},
  author={Sanyal, Paromita},
  journal={Qualitative Sociology},
  volume={38},
  number={2},
  pages={103--137},
  year={2015},
  publisher={Springer}
}

@article{berreman1972race,
  title={Race, caste, and other invidious distinctions in social stratification},
  author={Berreman, Gerald D},
  journal={Race},
  volume={13},
  number={4},
  pages={385--414},
  year={1972},
  publisher={Sage Publications Sage CA: Thousand Oaks, CA}
}

@article{patil2014caste,
  title={Caste-, work-, and descent-based discrimination as a determinant of health in social epidemiology},
  author={Patil, Rajan R},
  journal={Social work in public health},
  volume={29},
  number={4},
  pages={342--349},
  year={2014},
  publisher={Taylor \& Francis}
}

@article{christakis2010social,
  title={Social network sensors for early detection of contagious outbreaks},
  author={Christakis, Nicholas A and Fowler, James H},
  journal={PloS one},
  volume={5},
  number={9},
  pages={e12948},
  year={2010},
  publisher={Public Library of Science San Francisco, USA}
}

@article{hoff2002latent,
  title={Latent space approaches to social network analysis},
  author={Hoff, Peter D and Raftery, Adrian E and Handcock, Mark S},
  journal={Journal of the american Statistical association},
  volume={97},
  number={460},
  pages={1090--1098},
  year={2002},
  publisher={Taylor \& Francis}
}

@article{wang1987stochastic,
  title={Stochastic blockmodels for directed graphs},
  author={Wang, Yuchung J and Wong, George Y},
  journal={Journal of the American Statistical Association},
  volume={82},
  number={397},
  pages={8--19},
  year={1987},
  publisher={Taylor \& Francis}
}

@article{robins2007introduction,
  title={An introduction to exponential random graph (p*) models for social networks},
  author={Robins, Garry and Pattison, Pip and Kalish, Yuval and Lusher, Dean},
  journal={Social networks},
  volume={29},
  number={2},
  pages={173--191},
  year={2007},
  publisher={Elsevier}
}

@article{centola2010spread,
  title={The spread of behavior in an online social network experiment},
  author={Centola, Damon},
  journal={science},
  volume={329},
  number={5996},
  pages={1194--1197},
  year={2010},
  publisher={American Association for the Advancement of Science}
}

@article{smith2008social,
  title={Social networks and health},
  author={Smith, Kirsten P and Christakis, Nicholas A},
  journal={Annu. Rev. Sociol},
  volume={34},
  number={1},
  pages={405--429},
  year={2008},
  publisher={Annual Reviews}
}

@article{bokanyi2023anatomy,
  title={The anatomy of a population-scale social network},
  author={Bok{\'a}nyi, Eszter and Heemskerk, Eelke M and Takes, Frank W},
  journal={Scientific Reports},
  volume={13},
  number={1},
  pages={9209},
  year={2023},
  publisher={Nature Publishing Group UK London}
}

@book{putnam2000bowling,
  title={Bowling alone: The collapse and revival of American community},
  author={Putnam, Robert D},
  year={2000},
  publisher={Simon and schuster}
}

@article{harling2020age,
  title={Age and gender differences in social network composition and social support among older rural South Africans: Findings from the HAALSI study},
  author={Harling, Guy and Morris, Katherine Ann and Manderson, Lenore and Perkins, Jessica M and Berkman, Lisa F},
  journal={The Journals of Gerontology: Series B},
  volume={75},
  number={1},
  pages={148--159},
  year={2020},
  publisher={Oxford University Press US}
}

@article{carstensen1992social,
  title={Social and emotional patterns in adulthood: support for socioemotional selectivity theory.},
  author={Carstensen, Laura L},
  journal={Psychology and aging},
  volume={7},
  number={3},
  pages={331},
  year={1992},
  publisher={American Psychological Association}
}

@techreport{andrew2020mothers,
  title={Mothers’ social networks and socioeconomic gradients of isolation},
  author={Andrew, Alison and Attanasio, Orazio and Augsburg, Britta and Behrman, Jere and Day, Monimalika and Jervis, Pamela and Meghir, Costas and Phimister, Angus},
  year={2020},
  institution={National Bureau of Economic Research}
}

@article{perkins2015social,
  title={Social networks and health: a systematic review of sociocentric network studies in low-and middle-income countries},
  author={Perkins, Jessica M and Subramanian, SV and Christakis, Nicholas A},
  journal={Social science \& medicine},
  volume={125},
  pages={60--78},
  year={2015},
  publisher={Elsevier}
}

@article{lynch2022religiosity,
  title={Religiosity is associated with greater size, kin density, and geographic dispersal of women’s social networks in Bangladesh},
  author={Lynch, R and Schaffnit, S and Sear, R and Sosis, R and Shaver, J and Alam, N and Blumenfield, T and Mattison, SM and Shenk, M},
  journal={Scientific reports},
  volume={12},
  number={1},
  pages={18780},
  year={2022},
  publisher={Nature Publishing Group UK London}
}

@article{ghasemian2024structure,
  title={The structure and function of antagonistic ties in village social networks},
  author={Ghasemian, Amir and Christakis, Nicholas A},
  journal={Proceedings of the National Academy of Sciences},
  volume={121},
  number={26},
  pages={e2401257121},
  year={2024},
  publisher={National Academy of Sciences}
}

@article{airoldi2024induction,
  title={Induction of social contagion for diverse outcomes in structured experiments in isolated villages},
  author={Airoldi, Edoardo M and Christakis, Nicholas A},
  journal={Science},
  volume={384},
  number={6695},
  pages={eadi5147},
  year={2024},
  publisher={American Association for the Advancement of Science}
}

@article{shakya2019village,
  title={Do village-level normative and network factors help explain spatial variability in adolescent childbearing in rural Honduras?},
  author={Shakya, Holly B and Weeks, John R and Christakis, Nicholas A},
  journal={SSM-Population Health},
  volume={9},
  pages={100371},
  year={2019},
  publisher={Elsevier}
}

@book{gluckman1973judicial,
  title={The judicial process among the Barotse of Northern Rhodesia},
  author={Gluckman, Max},
  year={1973},
  publisher={Manchester University Press}
}

@article{dunbar1992neocortex,
  title={Neocortex size as a constraint on group size in primates},
  author={Dunbar, Robin IM},
  journal={Journal of human evolution},
  volume={22},
  number={6},
  pages={469--493},
  year={1992},
  publisher={Elsevier}
}

@article{woolcock1998social,
  title={Social capital and economic development: Toward a theoretical synthesis and policy framework},
  author={Woolcock, Michael},
  journal={Theory and society},
  volume={27},
  number={2},
  pages={151--208},
  year={1998},
  publisher={JSTOR}
}

@article{mcpherson2001birds,
  title={Birds of a feather: Homophily in social networks},
  author={McPherson, Miller and Smith-Lovin, Lynn and Cook, James M},
  journal={Annual review of sociology},
  volume={27},
  number={1},
  pages={415--444},
  year={2001},
  publisher={Annual Reviews 4139 El Camino Way, PO Box 10139, Palo Alto, CA 94303-0139, USA}
}

@article{harling2017leveraging,
  title={Leveraging contact network structure in the design of cluster randomized trials},
  author={Harling, Guy and Wang, Rui and Onnela, Jukka-Pekka and De Gruttola, Victor},
  journal={Clinical Trials},
  volume={14},
  number={1},
  pages={37--47},
  year={2017},
  publisher={SAGE Publications Sage UK: London, England}
}

\appendix

\section{Who is connected? - Tables}

\begin{table}[H]
\centering
\caption{Friendship degree (ZINB2, count part): estimates, standard errors, and $p$-values. Random intercepts for village, wave, and respondent.}
\label{tab:friend_est_se_p}
\small
\begin{tabular}{lrrr}
\toprule
Term & Estimate & SE & $p$ \\
\midrule
Intercept                       &  1.70900 & 0.02500 & $<10^{-100}$ \\
Health degree ($h_c$)           &  0.04500 & 0.00100 & $<10^{-16}$  \\
Financial degree ($fin_c$)      &  0.09600 & 0.00100 & $<10^{-16}$  \\
Adversarial degree ($adv_c$)    &  0.02200 & 0.00200 & $<10^{-16}$  \\
Age (centered) ($age_c$)        &  0.00380 & 0.00020 & $<10^{-16}$  \\
Education (centered) ($edu_c$)  &  0.00900 & 0.00150 & $1.2\times 10^{-9}$ \\
Male ($gender{=}1$)             & -0.03200 & 0.00680 & $2.0\times 10^{-6}$ \\
Income sufficiency (${=}1$)     & -0.01800 & 0.00560 & $1.1\times 10^{-3}$ \\
Married ($=1$)                  &  0.06200 & 0.00690 & $<10^{-16}$  \\
Religion: Protestant               &  0.05400 & 0.01100 & $9.4\times 10^{-7}$ \\
Religion: Catholic               &  0.03600 & 0.01100 & $1.6\times 10^{-3}$ \\
Indigenous ($=1$)               &  0.02600 & 0.01200 & $3.08\times 10^{-2}$ \\
Access to routes ($access_c$)   &  0.02500 & 0.02600 & $3.39\times 10^{-1}$ \\
Village size ($vsize_c$)        & -0.00096 & 0.00014 & $2.1\times 10^{-11}$ \\
\bottomrule
\end{tabular}
\normalsize
\end{table}

\begin{table}[H]
\centering
\caption{Health degree (ZINB2, count part; W1+W3 pooled): estimates, standard errors, and $p$-values. Random intercepts for village, wave, and respondent.}
\label{tab:health_est_se_p}
\small
\begin{tabular}{lrrr}
\toprule
Term & Estimate & SE & $p$ \\
\midrule
Intercept                         &  0.73670 & 0.03180 & $<10^{-100}$ \\
Friendship degree ($fr_c$)        &  0.05600 & 0.00100 & $<10^{-16}$  \\
Financial degree ($fin_c$)        &  0.07400 & 0.00190 & $<10^{-16}$  \\
Adversarial degree ($adv_c$)      &  0.03330 & 0.00310 & $<10^{-16}$  \\
Age (centered) ($age_c$)          &  0.00850 & 0.00030 & $<10^{-16}$  \\
Education (centered) ($edu_c$)    &  0.01400 & 0.00210 & $<10^{-10}$  \\
Male ($gender{=}1$)               & -0.37610 & 0.01010 & $<10^{-16}$  \\
Income sufficiency ($=1$)         & -0.02150 & 0.00840 & $1.0\times 10^{-2}$ \\
Married ($=1$)                    &  0.05880 & 0.01010 & $<10^{-7}$   \\
Religion: Protestant                 &  0.07860 & 0.01670 & $<10^{-5}$   \\
Religion: Catholic                 &  0.04960 & 0.01710 & $3.8\times 10^{-3}$ \\
Indigenous ($=1$)                 &  0.10210 & 0.01670 & $<10^{-8}$   \\
Access to routes ($access_c$)     & -0.00680 & 0.01450 & $6.37\times 10^{-1}$ \\
Village size ($vsize_c$)          & -0.00030 & 0.00010 & $5.9\times 10^{-2}$ \\
\bottomrule
\end{tabular}
\normalsize
\end{table}

\begin{table}[H]
\centering
\caption{Financial degree (ZINB2, count part); W1+W3 pooled): estimates, standard errors, and $p$-values. Random intercepts for village and respondent.}
\label{tab:financial_est_se_p}
\small
\begin{tabular}{lrrr}
\toprule
Term & Estimate & SE & $p$ \\
\midrule
Intercept                         &  0.47770 & 0.02340 & $7.5\times 10^{-93}$ \\
Friendship degree ($fr_c$)        &  0.07060 & 0.00088 & $<10^{-16}$  \\
Health degree ($h_c$)             &  0.03790 & 0.00128 & $<10^{-16}$  \\
Adversarial degree ($adv_c$)      &  0.03910 & 0.00306 & $<10^{-16}$  \\
Age (centered) ($age_c$)          &  0.00240 & 0.00033 & $1.75\times 10^{-13}$ \\
Education (centered) ($edu_c$)    &  0.01410 & 0.00207 & $1.01\times 10^{-11}$ \\
Male ($gender{=}1$)               &  0.21000 & 0.00938 & $<10^{-16}$  \\
Income sufficiency ($=1$)         &  0.04310 & 0.00837 & $2.64\times 10^{-7}$ \\
Married ($=1$)                    &  0.08000 & 0.00991 & $6.95\times 10^{-16}$ \\
Religion: Protestant                 &  0.06140 & 0.01580 & $9.79\times 10^{-5}$ \\
Religion: Catholic                 &  0.04910 & 0.01620 & $2.42\times 10^{-3}$ \\
Indigenous ($=1$)                 &  0.03930 & 0.01700 & $2.07\times 10^{-2}$ \\
Access to routes ($access_c$)     &  0.03260 & 0.02140 & $1.28\times 10^{-1}$ \\
Village size ($vsize_c$)          &  0.00006 & 0.00016 & $7.02\times 10^{-1}$ \\
\bottomrule
\end{tabular}
\normalsize
\end{table}

\begin{table}[H]
\centering
\caption{Adversarial degree (ZINB2, count part): estimates, standard errors, and $p$-values. Random intercepts for village, wave, and respondent.}
\label{tab:adversarial_est_se_p}
\small
\begin{tabular}{lrrr}
\toprule
Term & Estimate & SE & $p$ \\
\midrule
Intercept                         & -0.46900 & 0.15500 & $2.5\times 10^{-3}$ \\
Friendship degree ($fr_c$)        &  0.02130 & 0.00213 & $<2\times 10^{-16}$ \\
Financial degree ($fin_c$)        &  0.04280 & 0.00430 & $<2\times 10^{-16}$ \\
Health degree ($h_c$)             &  0.01920 & 0.00294 & $5.9\times 10^{-11}$ \\
Age (centered) ($age_c$)          & -0.00190 & 0.00078 & $1.46\times 10^{-2}$ \\
Education (centered) ($edu_c$)    & -0.02600 & 0.00442 & $4.3\times 10^{-9}$ \\
Male ($gender{=}1$)               & -0.46300 & 0.02460 & $<2\times 10^{-16}$ \\
Income sufficiency ($=1$)         & -0.04180 & 0.01770 & $1.8\times 10^{-2}$ \\
Married ($=1$)                    &  0.03720 & 0.02170 & $8.6\times 10^{-2}$ \\
Religion: Protestant                 & -0.00009 & 0.03370 & $9.98\times 10^{-1}$ \\
Religion: Catholic                 & -0.05300 & 0.03480 & $1.28\times 10^{-1}$ \\
Indigenous ($=1$)                 &  0.14200 & 0.03930 & $3.2\times 10^{-4}$ \\
Access to routes ($access_c$)     &  0.02530 & 0.05450 & $6.43\times 10^{-1}$ \\
Village size ($vsize_c$)          &  0.00052 & 0.00044 & $2.36\times 10^{-1}$ \\
\bottomrule
\end{tabular}
\normalsize
\end{table}

\noindent Tables~14--17 (ZINB2 count models with random intercepts) show a consistent picture of who is connected across the four layers. First, degrees are strongly \emph{multiplex}: having more ties in one layer is positively associated with having more ties in the other layers (e.g., friendship degree increases with health, financial, and adversarial degree; and adversarial degree also rises with cooperative degrees). Second, standard socio-demographic gradients are clear: age and education are positively associated with degree in all three cooperative layers (health, friendship, financial), but both are negatively associated with adversarial degree, suggesting that older and more educated individuals participate more in supportive networks while being less involved in conflict. Third, gender roles are layer-specific: men have fewer health and friendship ties, substantially more financial ties, and fewer adversarial ties. Fourth, being married is associated with higher degree in cooperative layers, while the association with adversarial degree is weaker and not statistically robust. Fifth, economic security shifts the type of connectedness: income sufficiency is associated with fewer health and friendship ties and fewer adversarial ties, but with more financial ties. Sixth, cultural identity variables matter mainly for cooperative ties: affiliation with the major religions is associated with higher cooperative degree, while religion is not a robust predictor of adversarial degree; indigenous identity is positively associated with degree in each layer, including adversarial ties. Finally, village/spatial covariates contribute little in these models: access to routes is not significant across layers, and village size is most clearly associated with lower friendship degree (with weaker or null associations in the other layers).

\section{Assortativity - Tables - Figures}

\begin{figure}[H]
  \centering
    \centering
    \scalebox{1.2}[1]{%
  \includegraphics[scale=0.5] {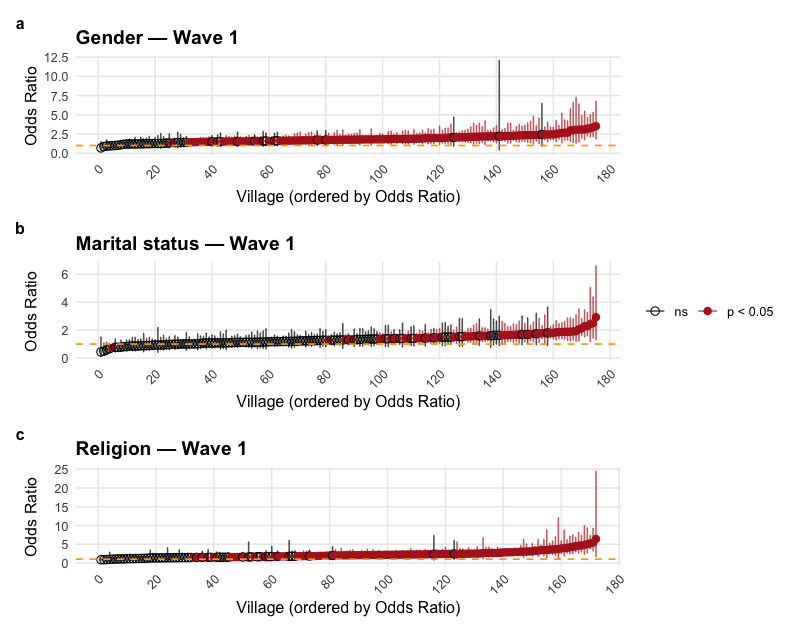}
    }
\caption{\textbf{Dyad-level assortativity in health ties (Wave 1).}
For each village, we fit dyad-level logistic regressions predicting the presence of a health tie from attribute similarity.
Panels report village-specific odds ratios (points) with 95\% confidence intervals for each attribute; villages are ordered by the odds ratio within each panel.
Red points indicate villages where the similarity coefficient is statistically significant ($p<0.05$).
The dashed horizontal line marks OR$=1$ (no association).
See Table~18 for corresponding numerical details.}
\label{fig:assort_health_w1}
\end{figure}

\begin{table}[H]
\centering
\caption{Health (Wave 1): degree‑aware permutation rejections (\# villages with $p<0.05$) by attribute and tolerance on group‑specific mean degrees ($\pm5\%$, $\pm10\%$, $\pm20\%$).}
\label{tab:health_w1_perm_counts}
\small
\begin{tabular}{llrrr}
\toprule
\textbf{Attribute} & \textbf{Category} & \textbf{$\pm$5\%} & \textbf{$\pm$10\%} & \textbf{$\pm$20\%} \\
\midrule
Gender              & M--M (assort.)                 & 53  & 136 & 131 \\
                    & F--F (assort.)                 & 56  & 111 &  45 \\
                    & M--F (dissort.)                &  0  &   0 &   0 \\
\addlinespace
Religion            & Catholic--Catholic (assort.)   & 101 &  94 &  66 \\
                    & Protestant--Protestant (assort.) & 100 & 101 &  92 \\
                    & Other--Other (assort.)         &  14 &  14 &  12 \\
                    & Different Religions (dissort.) &   0 &   0 &   0 \\
\addlinespace
Marital status      & Married--Married (assort.)     &   9 &  11 &  33 \\
                    & Single--Single (assort.)       &   6 &   6 &   3 \\
                    & Married--Single (dissort.)     &   4 &   4 &   3 \\
\addlinespace
Education           & High--High (assort.)           &  10 &   9 &  18 \\
                    & Low--Low (assort.)             &   4 &   2 &   3 \\
                    & High--Low (dissort.)           &   5 &   6 &   5 \\
\addlinespace
Age                 & Old--Old (assort.)             &   1 &   0 &   0 \\
                    & Young--Young (assort.)         &   2 &   6 &  32 \\
                    & Old--Young (dissort.)          &  22 &  20 &  15 \\
\addlinespace
Income sufficiency  & Sufficient--Sufficient (assort.)   &  0 &  0 &  0 \\
                    & Insufficient--Insufficient (assort.) & 0 &  0 &  0 \\
                    & Sufficient--Insufficient (dissort.) & 0 &  0 &  0 \\
\addlinespace
Indigenous (binary) & Indigenous--Indigenous (assort.)   & 16 & 16 & 17 \\
                    & Non-Indigenous--Non-Indigenous (assort.) & 15 & 9 & 5 \\
                    & Indigenous--Non-Indigenous (dissort.)    &  0 &  0 &  1 \\
\bottomrule
\end{tabular}
\normalsize
\end{table}

\begin{figure}[H]
  \centering
    \centering
  \scalebox{1.2}[1]{%
  \includegraphics[scale=0.5] {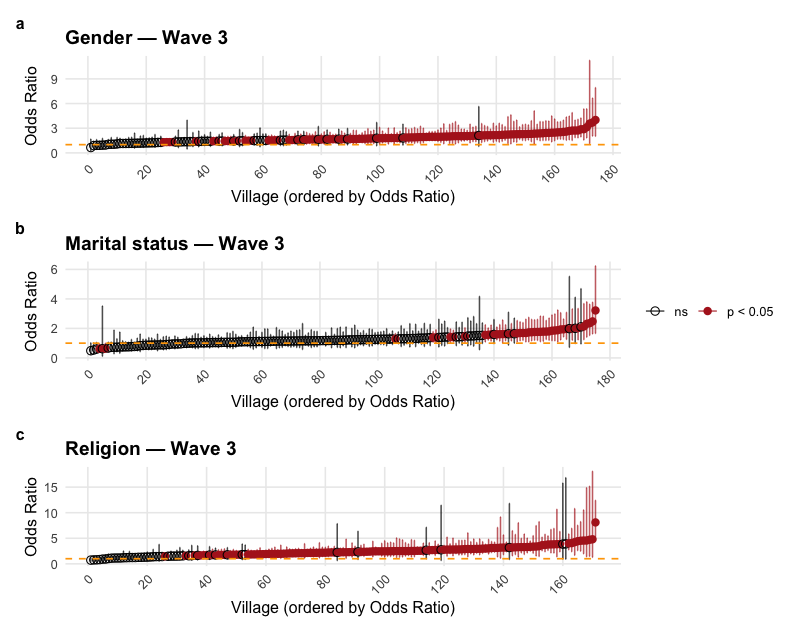}
    }
\caption{\textbf{Dyad-level assortativity in health ties (Wave 3).}
For each village, we fit dyad-level logistic regressions predicting the presence of a health tie from attribute similarity.
Panels report village-specific odds ratios (points) with 95\% confidence intervals for each attribute; villages are ordered by the odds ratio within each panel.
Red points indicate villages where the similarity coefficient is statistically significant ($p<0.05$).
The dashed horizontal line marks OR$=1$ (no association).
See Table~19 for corresponding numerical details.}
\label{fig:assort_health_w3}
\end{figure}

\begin{table}[H]
\centering
\caption{Health (Wave 3): degree‑aware permutation rejections (\# villages with $p<0.05$) by attribute and tolerance on group‑specific mean degrees ($\pm5\%$, $\pm10\%$, $\pm20\%$).}
\label{tab:health_w3_perm_counts}
\small
\begin{tabular}{llrrr}
\toprule
\textbf{Attribute} & \textbf{Category} & \textbf{$\pm$5\%} & \textbf{$\pm$10\%} & \textbf{$\pm$20\%} \\
\midrule
Gender              & M--M (assort.)                 & 53  & 136 & 131 \\
                    & F--F (assort.)                 & 56  & 111 &  45 \\
                    & M--F (dissort.)                &  0  &   0 &   0 \\
\addlinespace
Religion            & Catholic--Catholic (assort.)   & 101 &  94 &  66 \\
                    & Protestant--Protestant (assort.) & 100 & 101 &  92 \\
                    & Other--Other (assort.)         &  14 &  14 &  12 \\
                    & Different Religions (dissort.) &   0 &   0 &   0 \\
\addlinespace
Marital status      & Married--Married (assort.)     &   9 &  11 &  33 \\
                    & Single--Single (assort.)       &   6 &   6 &   3 \\
                    & Married--Single (dissort.)     &   4 &   4 &   3 \\
\addlinespace
Education           & High--High (assort.)           &  10 &   9 &  18 \\
                    & Low--Low (assort.)             &   4 &   2 &   3 \\
                    & High--Low (dissort.)           &   5 &   6 &   5 \\
\addlinespace
Age                 & Old--Old (assort.)             &   1 &   0 &   0 \\
                    & Young--Young (assort.)         &   2 &   6 &  32 \\
                    & Old--Young (dissort.)          &  22 &  20 &  15 \\
\addlinespace
Income sufficiency  & Sufficient--Sufficient (assort.)   &  0 &  0 &  0 \\
                    & Insufficient--Insufficient (assort.) & 0 &  0 &  0 \\
                    & Sufficient--Insufficient (dissort.) & 0 &  0 &  0 \\
\addlinespace
Indigenous (binary) & Indigenous--Indigenous (assort.)   & 16 & 16 & 17 \\
                    & Non-Indigenous--Non-Indigenous (assort.) & 15 & 9 & 5 \\
                    & Indigenous--Non-Indigenous (dissort.)    &  0 &  0 &  1 \\
\bottomrule
\end{tabular}
\normalsize
\end{table}

  \begin{figure}[H]
    \centering
  \scalebox{1.2}[1]{%
  \includegraphics[scale=0.5] {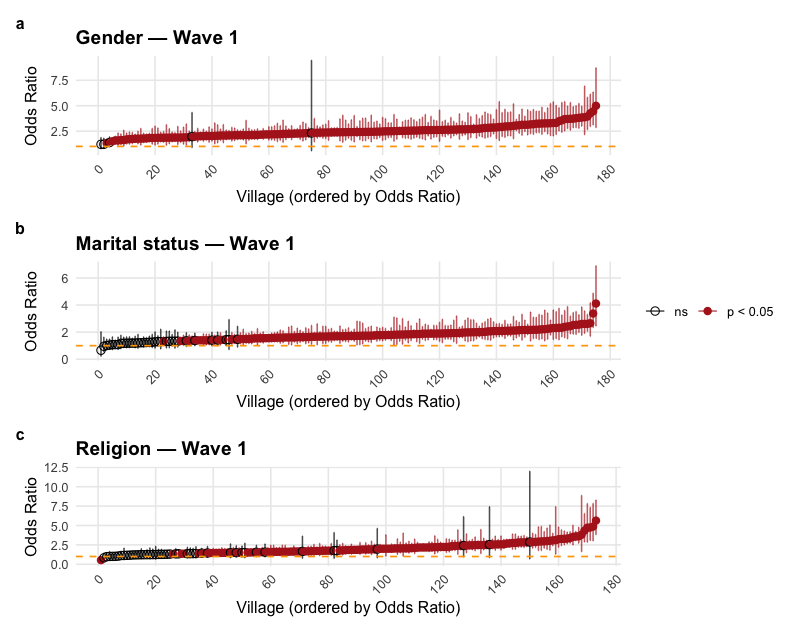}
    }
\caption{\textbf{Dyad-level assortativity in friendship ties (Wave 1).}
For each village, we fit dyad-level logistic regressions predicting the presence of a friendship tie from attribute similarity.
Panels report village-specific odds ratios (points) with 95\% confidence intervals for each attribute; villages are ordered by the odds ratio within each panel.
Red points indicate villages where the similarity coefficient is statistically significant ($p<0.05$).
The dashed horizontal line marks OR$=1$ (no association).
See Table~20 for corresponding numerical details.}
\label{fig:assort_friendship_w1}
\end{figure}

\begin{table}[H]
\centering
\caption{Friendship (Wave 1): degree-aware permutation rejections (\# villages with $p<0.05$) by attribute and tolerance.}
\label{tab:friendship_w1_perm_counts}
\small
\begin{tabular}{llrrr}
\toprule
\textbf{Attribute} & \textbf{Category} & \textbf{$\pm$5\%} & \textbf{$\pm$10\%} & \textbf{$\pm$20\%} \\
\midrule
Gender                 & M--M                 & 170 & 0   & 0   \\
                       & F--F                 & 168 & 0   & 0   \\
                       & M--F                 & 0   & 0   & 0   \\
\addlinespace
Religion               & Catholic -- Catholic & 105 & 96  & 85  \\
                       & Protestant -- Protestant & 118 & 112 & 105 \\
                       & Non -- Non           & 31  & 27  & 22  \\
                       & Different Religions  & 0   & 1   & 1   \\
\addlinespace
Marital status         & Not Single--Not Single & 120 & 118 & 130 \\
                       & Single--Single       & 117 & 95  & 51  \\
                       & Not Single--Single   & 0   & 0   & 0   \\
\addlinespace
Education              & High--High           & 83  & 88  & 84  \\
                       & Low--Low             & 64  & 30  & 19  \\
                       & High--Low            & 0   & 0   & 0   \\
\addlinespace
Age                    & Old--Old             & 93  & 87  & 17  \\
                       & Young--Young         & 112 & 141 & 167 \\
                       & Old--Young           & 0   & 0   & 0   \\
\addlinespace
Income sufficiency     & Sufficient -- Sufficient & 34 & 17 & 17 \\
                       & Non--Non             & 44  & 36  & 35  \\
                       & Sufficient -- Non    & 1   & 2   & 1   \\
\addlinespace
Indigenous (binary)    & Indig.--Indig.       & 28  & 32  & 32  \\
                       & Non--Non             & 22  & 17  & 8   \\
                       & Indig.--Non          & 1   & 0   & 0   \\
\bottomrule
\end{tabular}
\end{table}

  \begin{figure}[H]
    \centering
  \scalebox{1.2}[1]{%
  \includegraphics[scale=0.5] {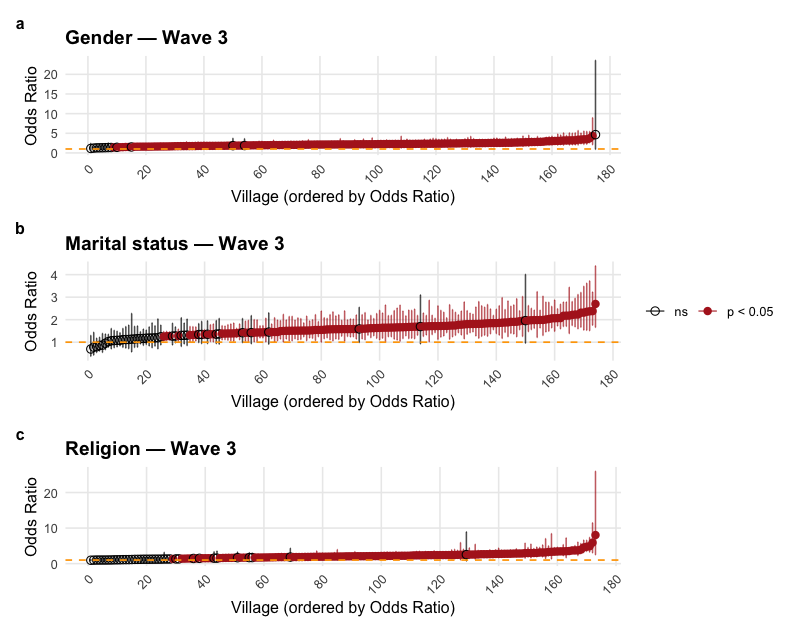}
    }
\caption{\textbf{Dyad-level assortativity in friendship ties (Wave 3).}
For each village, we fit dyad-level logistic regressions predicting the presence of a friendship tie from attribute similarity.
Panels report village-specific odds ratios (points) with 95\% confidence intervals for each attribute; villages are ordered by the odds ratio within each panel.
Red points indicate villages where the similarity coefficient is statistically significant ($p<0.05$).
The dashed horizontal line marks OR$=1$ (no association).
See Table~21 for corresponding numerical details.}
\label{fig:assort_friendship_w3}
\end{figure}

\begin{table}[H]
\centering
\caption{Friendship (Wave 3): degree-aware permutation rejections (\# villages with $p<0.05$) by attribute and tolerance.}
\label{tab:friendship_w3_perm_counts}
\small
\begin{tabular}{llrrr}
\toprule
\textbf{Attribute} & \textbf{Category} & \textbf{$\pm$5\%} & \textbf{$\pm$10\%} & \textbf{$\pm$20\%} \\
\midrule
Gender               & M--M                      & 53 & 136 & 131 \\
                     & F--F                      & 56 & 111 & 45 \\
                     & M--F                      & 0 & 0 & 0 \\
\addlinespace
Religion             & Catholic -- Catholic      & 109 & 102 & 89 \\
                     & Protestant -- Protestant  & 118 & 112 & 104 \\
                     & Non -- Non                & 27 & 26 & 18 \\
                     & Different Religions       & 0 & 0 & 0 \\
\addlinespace
Marital status       & Not Single--Not Single    & 82 & 92 & 112 \\
                     & Single--Single            & 67 & 50 & 18 \\
                     & Not Single--Single        & 0 & 0 & 0 \\
\addlinespace
Education            & High--High                & 66 & 63 & 65 \\
                     & Low--Low                  & 38 & 19 & 13 \\
                     & High--Low                 & 0 & 0 & 0 \\
\addlinespace
Age                  & Old--Old                  & 57 & 37 & 3 \\
                     & Young--Young              & 87 & 123 & 146 \\
                     & Old--Young                & 0 & 0 & 0 \\
\addlinespace
Income sufficiency   & Sufficient -- Sufficient  & 0 & 0 & 0 \\
                     & Non--Non                  & 0 & 0 & 0 \\
                     & Sufficient -- Non         & 0 & 0 & 0 \\
\addlinespace
Indigenous (binary)  & Indig.--Indig.            & 23 & 26 & 30 \\
                     & Non--Non                  & 20 & 15 & 6 \\
                     & Indig.--Non               & 0 & 1 & 1 \\
\bottomrule
\end{tabular}
\end{table}

  \begin{figure}[H]
    \centering
     \scalebox{1.2}[1]{%
  \includegraphics[scale=0.5] {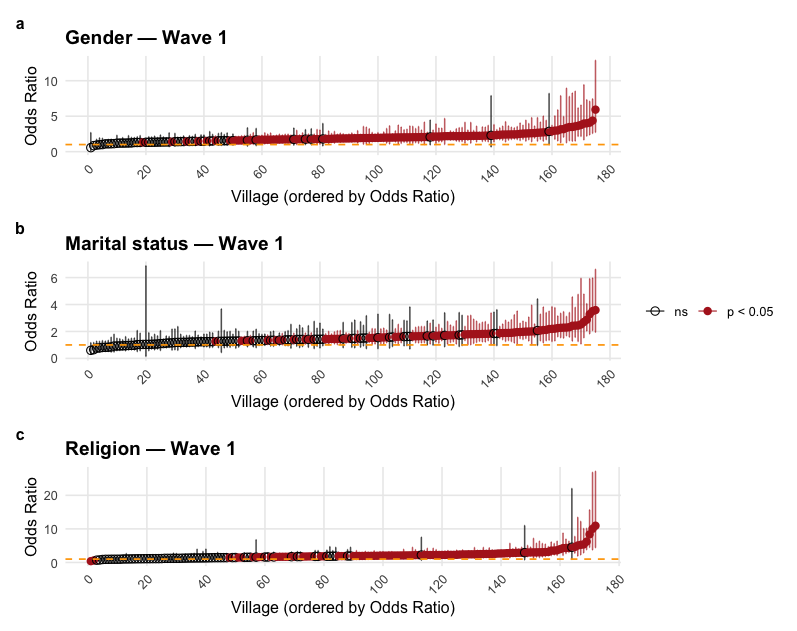}
  }
\caption{\textbf{Dyad-level assortativity in financial-help ties (Wave 1).}
For each village, we fit dyad-level logistic regressions predicting the presence of a financial-help tie from attribute similarity.
Panels report village-specific odds ratios (points) with 95\% confidence intervals for each attribute; villages are ordered by the odds ratio within each panel.
Red points indicate villages where the similarity coefficient is statistically significant ($p<0.05$).
The dashed horizontal line marks OR$=1$ (no association).
See Table~22 for corresponding numerical details.}
\label{fig:assort_financial_w1}
\end{figure}


\begin{table}[H]
\centering
\caption{Financial (Wave 1): degree-aware permutation rejections (\# villages with $p<0.05$) by attribute and tolerance.}
\label{tab:financial_w1_perm_counts}
\small
\begin{tabular}{llrrr}
\toprule
\textbf{Attribute} & \textbf{Category} & \textbf{$\pm$5\%} & \textbf{$\pm$10\%} & \textbf{$\pm$20\%} \\
\midrule
Gender               & M--M                      & 53 & 136 & 131 \\
                     & F--F                      & 56 & 111 & 45 \\
                     & M--F                      & 0 & 0 & 0 \\
\addlinespace
Religion             & Catholic -- Catholic      & 95 & 91 & 76 \\
                     & Protestant -- Protestant  & 99 & 93 & 81 \\
                     & Non -- Non                & 7 & 7 & 6 \\
                     & Different Religions       & 1 & 1 & 1 \\
\addlinespace
Marital status       & Not Single--Not Single    & 58 & 57 & 65 \\
                     & Single--Single            & 52 & 40 & 18 \\
                     & Not Single--Single        & 0 & 0 & 0 \\
\addlinespace
Education            & High--High                & 38 & 31 & 30 \\
                     & Low--Low                  & 27 & 14 & 9 \\
                     & High--Low                 & 2 & 1 & 0 \\
\addlinespace
Age                  & Old--Old                  & 40 & 24 & 4 \\
                     & Young--Young              & 58 & 72 & 101 \\
                     & Old--Young                & 0 & 0 & 0 \\
\addlinespace
Income sufficiency   & Sufficient -- Sufficient  & 26 & 22 & 13 \\
                     & Non--Non                  & 26 & 20 & 20 \\
                     & Sufficient -- Non         & 4 & 3 & 3 \\
\addlinespace
Indigenous (binary)  & Indig.--Indig.            & 17 & 20 & 19 \\
                     & Non--Non                  & 13 & 12 & 6 \\
                     & Indig.--Non               & 0 & 0 & 0 \\
\bottomrule
\end{tabular}
\end{table}

  \begin{figure}[H]
    \centering
     \scalebox{1.2}[1]{%
  \includegraphics[scale=0.5] {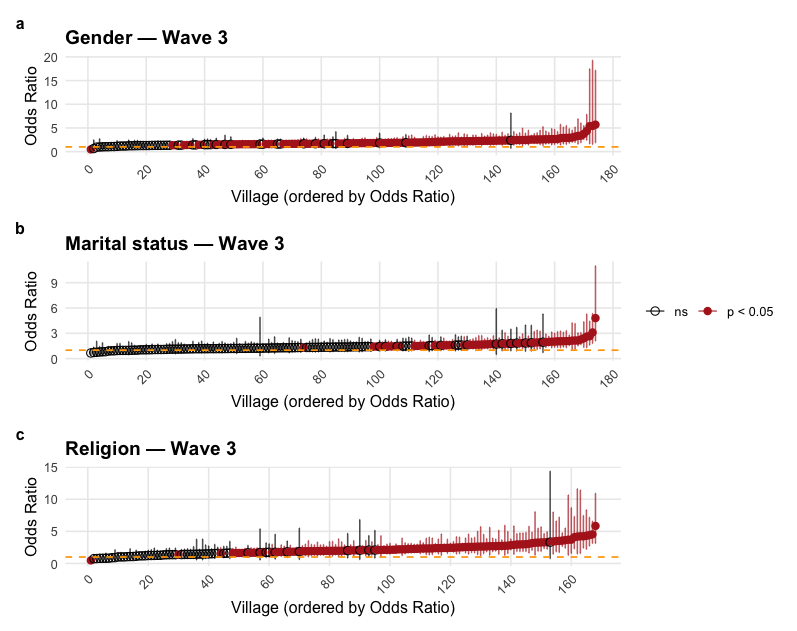}
    }
\caption{\textbf{Dyad-level assortativity in financial-help ties (Wave 3).}
For each village, we fit dyad-level logistic regressions predicting the presence of a financial-help tie from attribute similarity.
Panels report village-specific odds ratios (points) with 95\% confidence intervals for each attribute; villages are ordered by the odds ratio within each panel.
Red points indicate villages where the similarity coefficient is statistically significant ($p<0.05$).
The dashed horizontal line marks OR$=1$ (no association).
See Table~23 for corresponding numerical details.}
\label{fig:assort_financial_w3}
\end{figure}

\begin{table}[H]
\centering
\caption{Financial (Wave 3): degree-aware permutation rejections (\# villages with $p<0.05$) by attribute and tolerance.}
\label{tab:financial_w3_perm_counts}
\small
\begin{tabular}{llrrr}
\toprule
\textbf{Attribute} & \textbf{Category} & \textbf{$\pm$5\%} & \textbf{$\pm$10\%} & \textbf{$\pm$20\%} \\
\midrule
Gender               & M--M & 53 & 136 & 131 \\
                    & F--F & 56 & 111 & 45 \\
                    & M--F & 0 & 0 & 0 \\
\addlinespace
Religion             & Catholic -- Catholic & 95 & 90 & 81 \\
                    & Protestant -- Protestant & 98 & 92 & 84 \\
                    & Non -- Non & 15 & 12 & 13 \\
                    & Different Religions & 0 & 0 & 1 \\
\addlinespace
Marital status       & Not Single--Not Single & 33 & 34 & 55 \\
                    & Single--Single & 27 & 24 & 10 \\
                    & Not Single--Single & 1 & 0 & 0 \\
\addlinespace
Education            & High--High & 27 & 21 & 18 \\
                    & Low--Low & 23 & 10 & 7 \\
                    & High--Low & 3 & 1 & 4 \\
\addlinespace
Age                  & Old--Old & 25 & 7 & 1 \\
                    & Young--Young & 36 & 51 & 79 \\
                    & Old--Young & 3 & 1 & 0 \\
\addlinespace
Income sufficiency   & Sufficient -- Sufficient & 0 & 0 & 0 \\
                    & Non--Non & 0 & 0 & 0 \\
                    & Sufficient -- Non & 0 & 0 & 0 \\
\addlinespace
Indigenous (binary)  & Indig.--Indig. & 19 & 16 & 16 \\
                    & Non--Non & 14 & 11 & 5 \\
                    & Indig.--Non & 0 & 1 & 1 \\
\bottomrule
\end{tabular}
\end{table}

  \begin{figure}[H]
    \centering
    \scalebox{1.2}[1]{%
  \includegraphics[scale=0.5] {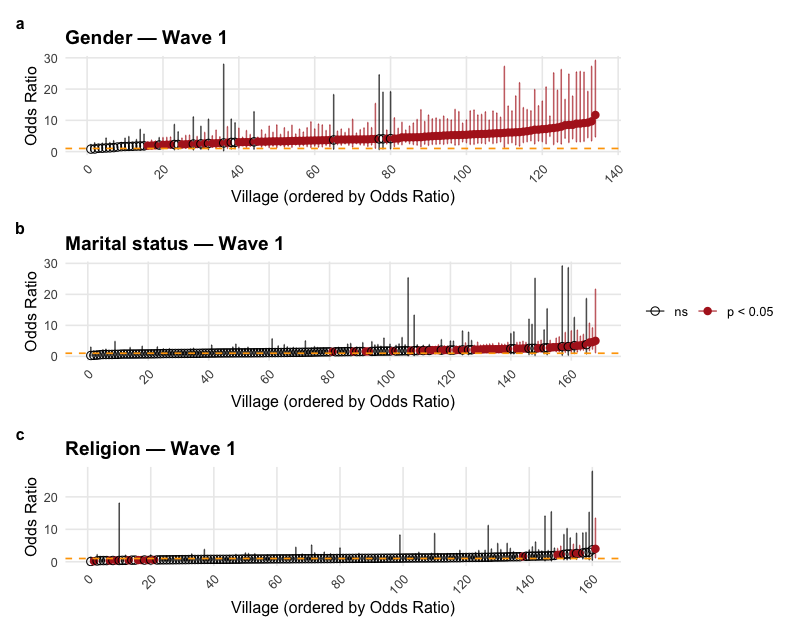}
    }
\caption{\textbf{Dyad-level assortativity in adversarial ties (Wave 1).}
For each village, we fit dyad-level logistic regressions predicting the presence of an adversarial tie from attribute similarity.
Panels report village-specific odds ratios (points) with 95\% confidence intervals for each attribute; villages are ordered by the odds ratio within each panel.
Red points indicate villages where the similarity coefficient is statistically significant ($p<0.05$).
The dashed horizontal line marks OR$=1$ (no association).
See Tables~24 for corresponding numerical details.}
\label{fig:assort_adversarial_w1}
\end{figure}

\begin{table}[H]
\centering
\caption{Adversarial 1: degree-aware permutation rejections (\# villages with $p<0.05$) by attribute and tolerance.}
\label{tab:adversarial1_perm_counts}
\small
\begin{tabular}{llrrr}
\toprule
\textbf{Attribute} & \textbf{Category} & \textbf{$\pm$5\%} & \textbf{$\pm$10\%} & \textbf{$\pm$20\%} \\
\midrule
Gender              & M--M            & 76 & 136 & 131 \\
                    & F--F            & 77 & 111 & 45  \\
                    & M--F            & 0  & 0   & 0   \\
\addlinespace
Religion            & Catholic -- Catholic       & 6  & 5  & 7 \\
                    & Protestant -- Protestant   & 4  & 4  & 2 \\
                    & Non -- Non                 & 1  & 0  & 1 \\
                    & Different Religions        & 4  & 6  & 7 \\
\addlinespace
Marital status      & Not Single--Not Single     & 34 & 33 & 28 \\
                    & Single--Single             & 36 & 35 & 25 \\
                    & Not Single--Single         & 1  & 0  & 0  \\
\addlinespace
Education           & High--High                 & 15 & 18 & 14 \\
                    & Low--Low                   & 14 & 10 & 3  \\
                    & High--Low                  & 2  & 3  & 3  \\
\addlinespace
Age                 & Old--Old                   & 52 & 34 & 20 \\
                    & Young--Young               & 54 & 54 & 59 \\
                    & Old--Young                 & 0  & 0  & 0  \\
\addlinespace
Income sufficiency  & Sufficient -- Sufficient   & 1  & 1  & 0  \\
                    & Non--Non                   & 2  & 2  & 1  \\
                    & Sufficient -- Non          & 4  & 5  & 5  \\
\addlinespace
Indigenous (binary) & Indig.--Indig.             & 3  & 3  & 4  \\
                    & Non--Non                   & 4  & 4  & 2  \\
                    & Indig.--Non                & 0  & 0  & 0  \\
\bottomrule
\end{tabular}
\end{table}

  \begin{figure}[H]
    \centering
     \scalebox{1.2}[1]{%
  \includegraphics[scale=0.5] {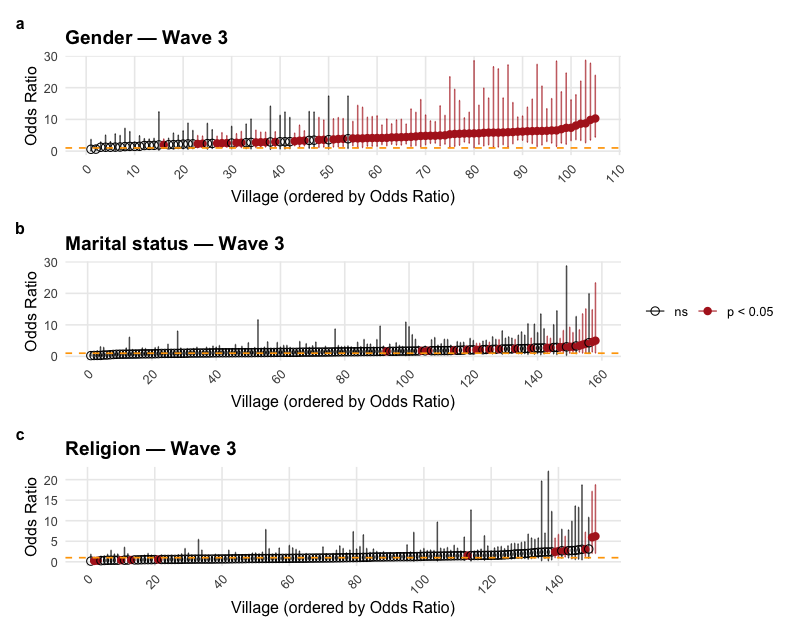}
    }
\caption{\textbf{Dyad-level assortativity in adversarial ties (Wave 3).}
For each village, we fit dyad-level logistic regressions predicting the presence of an adversarial tie from attribute similarity.
Panels report village-specific odds ratios (points) with 95\% confidence intervals for each attribute; villages are ordered by the odds ratio within each panel.
Red points indicate villages where the similarity coefficient is statistically significant ($p<0.05$).
The dashed horizontal line marks OR$=1$ (no association).
See Table~25 for corresponding numerical details.}
\label{fig:assort_adversarial_w3}
\end{figure}

\begin{table}[H]
\centering
\caption{Adversarial 3: degree-aware permutation rejections (\# villages with $p<0.05$) by attribute and tolerance.}
\label{tab:w3_perm_counts}
\small
\begin{tabular}{llrrr}
\toprule
\textbf{Attribute} & \textbf{Category} & \textbf{$\pm$5\%} & \textbf{$\pm$10\%} & \textbf{$\pm$20\%} \\
\midrule
Gender              & M--M            & 53 & 136 & 131 \\
                    & F--F            & 56 & 111 & 45 \\
                    & M--F            & 0  & 0   & 0  \\
\addlinespace
Religion            & Catholic -- Catholic         & 8  & 7   & 5  \\
                    & Protestant -- Protestant     & 7  & 7   & 6  \\
                    & Non -- Non                   & 3  & 3   & 2  \\
                    & Different Religions          & 3  & 4   & 5  \\
\addlinespace
Marital status      & Not Single--Not Single       & 16 & 10  & 4  \\
                    & Single--Single               & 16 & 10  & 8  \\
                    & Not Single--Single           & 0  & 0   & 1  \\
\addlinespace
Education           & High--High                   & 9  & 7   & 11 \\
                    & Low--Low                     & 8  & 7   & 1  \\
                    & High--Low                    & 1  & 1   & 2  \\
\addlinespace
Age                 & Old--Old                     & 15 & 13  & 4  \\
                    & Young--Young                 & 15 & 16  & 15 \\
                    & Old--Young                   & 2  & 1   & 2  \\
\addlinespace
Income sufficiency  & Sufficient -- Sufficient     & 5  & 4   & 1  \\
                    & Non--Non                     & 5  & 5   & 4  \\
                    & Sufficient -- Non            & 3  & 3   & 4  \\
\addlinespace
Indigenous (binary) & Indig.--Indig.               & 3  & 2   & 3  \\
                    & Non--Non                     & 2  & 1   & 1  \\
                    & Indig.--Non                  & 0  & 0   & 0  \\
\bottomrule
\end{tabular}
\end{table}\label{fig:dyad_or_layers}

\section{Intervention and Data Collection Timeline}
\label{app:timeline}

\subsection*{Study Region}
The study region covered more than 200 square miles of mountainous terrain near the Guatemala border, with an estimated population of roughly 92,000 residents. Out of 238 identified towns and villages, 176 were selected on the basis of population size, accessibility, and safety. The adult population in these study villages was estimated at 32,800 individuals.\\

\noindent All persons aged 12 years or older, living or working in the study villages, were eligible at baseline. Individuals unable to provide consent due to cognitive impairment were excluded.

\subsection*{Recruitment}
Prior to launching fieldwork, we mapped settlements in the region to capture terrain, rainfall patterns, and distances to health facilities. A photographic census was then conducted using tablet-based software. Photos, GPS coordinates, and demographic details (age, sex, marital status) were recorded for all residents. Out of an estimated 32,800 eligible individuals, 93\% (30,422) consented to be included in the census. Village-level participation ranged from 55 to 627 respondents, with households averaging 2.8 participants.\\

\noindent Key descriptive characteristics were as follows: mean village size was 173 individuals (range 55–627), mean household size was 2.8 (range 1–13), and mean age was 33 (range 12–94). Fifty-four percent were women and 58\% reported being married or living as married.

\begin{table}[H]
\centering
\caption{Baseline Census Demographics, $N=30{,}422$}
\label{tab:census_demo}
\begin{tabular}{lcc}
\toprule
Characteristic & Mean & Range \\
\midrule
Village size & 173 & [55 -- 627] \\
Household size & 2.8 & [1 -- 13] \\
Age (years) & 33 & [12 -- 94] \\
Women & 54\% & --- \\
Married / living as married & 58\% & --- \\
\bottomrule
\end{tabular}
\end{table}

\begin{table}[H]
\centering
\caption{Baseline Cohort Characteristics, $N=24{,}812$}
\label{tab:cohort_stats}
\begin{tabular}{lc}
\toprule
Characteristic & Value  \\
\midrule
Age (years) & 33  [12 -- 93] \\
Women & 58\%  \\
Married / living as married & 58\% \\
Less than primary education & 70\% \\
Religion: Catholic & 51\% \\
Religion: Protestant & 32\%  \\
No religion & 16\%  \\
Indigenous & 12\%  \\
Self-reported general health fair/poor & 44\%  \\
Self-reported mental health fair/poor & 40\%  \\
\bottomrule
\end{tabular}
\end{table}

\subsection*{Timeline of Data Collection Waves}

We describe the timeline of the data collection and each wave:

\begin{itemize}
    \item Wave 0 (Jun 2015–Dec 2015): Photographic census of 176 villages; 93\% coverage of residents aged 12+ ($N = 30{,}422$).
    \item Wave 1 (Oct 2015–Jun 2016): Baseline survey with sociocentric network mapping, health attitudes, and behaviors ($N \approx 24{,}700$).
    \item Wave 2 (Jan 2018–Aug 2018): Interim survey 12 months into intervention; updated locations and statuses ($N \approx 21{,}500$).
    \item Wave 3 (Jan 2019–Dec 2019): New census of adults 15+, followed by endline survey with networks, health, and behaviors ($N \approx 22{,}500$).

\end{itemize}

\subsection*{Field Operations}
More than 100 local enumerators were trained to (i) build infrastructure and conduct preliminary data collection, (ii) recruit participants and conduct census enumeration, and (iii) administer surveys and interviews. Four field offices were established in strategic locations to minimize travel time. Each office was equipped with tablets, netbooks, printers, high-speed internet, and local servers for secure data transfer and synchronization. Enumerators received intensive training on instruments and software by U.S.-based project managers and were supervised daily by Honduran coordinators.

\subsection*{Community Engagement}
Partnerships were developed with municipal and health authorities, and meetings were held with health staff and community health workers. The Ministry of Health reviewed and approved all study protocols, consent processes, and instruments. Village leaders and indigenous councils were engaged in advance to present study objectives before recruitment and data collection.

\subsection*{Missing Data and Attrition}
Among the $\sim$32,800 eligible individuals, 93\% consented to participate, reducing baseline non-response bias. We attempted to recontact and track participants who moved within the region during follow-ups. Of 5,633 randomized households, 4,861 (86\%) had at least one respondent complete the endline survey. Attrition did not differ by treatment status ($\chi^2(1, N=5633) = 0.80$, $p = 0.37$).

\section{Inclusion Criteria}
Respondents were included in analyses if they met the following criteria: (a) they completed all applicable survey forms, (b) they were enrolled in the census and participated in the questionnaire of either Wave 1 and Wave 3 surveys.

\end{document}